\documentclass[pre,aps,preprint,10pt,showpacs,amsmath,amssymb,showkeys,floatfix,nofootinbib]{revtex4-1}
\usepackage{graphicx}
\usepackage{nicefrac}

\newcommand{\aap}{Astron. Astrophys.}  
\newcommand{\mnras}{Mon.\ Not.\ R.\ Astron.\ Soc.}  
\newcommand{\gafd}{Geophys.\ Astrophys.\ Fluid\ Dyn.}  
\newcommand{\apss}{Astrophysics and Space Science}  
\renewcommand{\vec}[1]{\mbox{\boldmath$#1$}}

\def\ga{\mathrel{\mathchoice {\vcenter{\offinterlineskip\halign{\hfil
$\displaystyle##$\hfil\cr>\cr\sim\cr}}}
{\vcenter{\offinterlineskip\halign{\hfil$\textstyle##$\hfil\cr
>\cr\sim\cr}}}
{\vcenter{\offinterlineskip\halign{\hfil$\scriptstyle##$\hfil\cr
>\cr\sim\cr}}}
{\vcenter{\offinterlineskip\halign{\hfil$\scriptscriptstyle##$\hfil\cr
>\cr\sim\cr}}}}}

\begin{document}

\preprint{APS/123-QED}

\title{Impact of time-dependent non-axisymmetric velocity perturbations\\ 
on dynamo action of von-K\'arm\'an-like flows.}

\author{Andr\'e Giesecke}
\email{a.giesecke@hzdr.de}

\author{Frank Stefani}
\affiliation{
Institute of Fluid Dynamics,\\
Helmholtz-Zentrum Dresden-Rossendorf\\
P. O. Box 51 01 19, D-01314 Dresden, Germany
}
\author{Javier Burguete}
\affiliation{
Departamento de F\'isica y Matem\'atica Aplicada,\\
Universidad de Navarra,\\
Irunlarrea 1, 31008 Pamplona, Spain
}

\date{\today}

\begin{abstract}
We present numerical simulations of the kinematic induction
equation in order to examine the dynamo efficiency of an axisymmetric
von K{\'a}rm{\'a}n--like flow subject to time-dependent
nonaxisymmetric velocity perturbations.  The numerical model is based
on the setup of the French von K{\'a}rm{\'a}n-sodium dynamo (VKS) and
on the flow measurements from a water experiment conducted at
the University of Navarra in Pamplona, Spain.
The principal experimental observations that are modeled
in our simulations are nonaxisymmetric vortexlike structures which
perform an azimuthal drift motion in the equatorial plane.
Our simulations show that the interactions of these periodic flow
perturbations with the fundamental drift of the magnetic 
eigenmode (including the special case of nondrifting fields)
essentially determine the temporal behavior of the dynamo state. 

We find two distinct regimes of dynamo action that
depend on the (prescribed) drift frequency of an ($m=2$) vortexlike
flow perturbation. For comparatively slowly drifting vortices we
observe a narrow window with enhanced growth rates and a drift of the
magnetic eigenmode that is synchronized with the perturbation
drift. The resonance-like enhancement of the growth rates takes place
when the vortex drift frequency roughly equals the drift frequency of
the magnetic eigenmode in the unperturbed system. Outside of this
small window, the field generation is hampered compared to the
unperturbed case, and the field amplitude of the magnetic eigenmode is
modulated with approximately twice the vortex drift frequency. The
abrupt transition between the resonant regime and the modulated regime
is identified as a spectral exceptional point where eigenvalues
(growth-rates and frequencies) and eigenfunctions of two previously
independent modes collapse.

In the actual configuration the drift frequencies
of the velocity perturbations that are observed in the water
experiment are much larger than the fundamental drift frequency of the
magnetic eigenmode that is obtained from our numerical
simulations. 
Hence, we conclude that the
fulfillment of the resonance condition might be unlikely in present
day dynamo experiments.  
However, a possibility to increase the dynamo efficiency
in the VKS experiment  might be
realized by an application of holes or fingers 
on the outer boundary in the equatorial plane. These mechanical
distortions provoke an anchorage of the vortices at fixed
positions thus allowing an adjustment of the temporal
behavior of the nonaxisymmetric flow perturbations.
\end{abstract}

\maketitle

\section{Introduction}
Cosmic magnetic fields are ubiquitous phenomena that are intrinsically
coupled to most astrophysical objects like planets, stars, or galaxies.
The origin of these fields involves the formation of electrical
currents by means of a complex flow of a conducting fluid or plasma.
This process, the so-called dynamo effect, is necessarily three
dimensional and nonlinear, which makes an analytical or numerical
approach difficult.  Meanwhile, fluid-flow-driven generation of
magnetic fields has also been obtained in laboratory experiments
providing a complementary tool to astronomical observations or
numerical simulations.  However, whereas astrophysical dynamo action
is comparably easy because of the large dimensions of the involved
flows, its experimental realization requires considerable technical
efforts \cite{ISI:000262272000001}.  An important obstacle for the
occurrence of laboratory dynamo action arises from the scaling
behavior of the power that is required to drive a flow with a
requested magnetic Reynolds number, ${\rm{Rm}}$.  For turbulent flows
this power scales $\propto {\rm{Rm}}^3$ so a reduction of the
critical ${\rm{Rm}}$ for the onset of dynamo action is most important
to achieve magnetic self-excitation at all.

So far, dynamo experiments based on a flow of a conducting fluid have
been successfully conducted in Riga \citep{2000PhRvL..84.4365G},
Karlsruhe \citep{abs}, and Cadarache \citep{2007PhRvL..98d4502M}.  The
first two facilities made use of a more or less predetermined fluid
flow essentially fixed by the forcing and the shape of the internal
tubes.  Note, however, that, at least in the Riga dynamo experiment, the
saturation process involved a nontrivial back-reaction effect
of the magnetic field that changes the geometry of the flow. Such
effects might be even more pronounced in the Cadarache
von-K{\'a}rm{\'a}n-sodium (VKS) dynamo. In that experiment, the flow
driving by two opposing impellers provides more freedom for the
development of a saturated turbulent state, in which the back-reaction
of the magnetic field on the fluid can strongly modify the geometry
and dynamics of the flow.  In an idealizing model the mean
axisymmetric flow between counter-rotating impellers comprises two
toroidal and two poloidal eddies (so-called {\it{s2t2}} topology), and
it is well known that this flow is able to drive a dynamo
\cite{1989RSPSA.425..407D,1998PhRvE..58.7397O}.  Various attempts in
different geometries have been made (numerically as well as
experimentally) in order to examine dynamo action driven by such a
flow
\citep{forest2002,2003EPJB...33..469M,2004mag...bourg,2005physics..11149S,
  2005PhFl...17k7104R,2007EL.....7759001B,2007PhRvL..98d4502M,2008EL.....8229001G,
  2008PhRvL.101n4502G,2009NJPh...11a3027R,2009PhRvE..80e6304R,2009PhFl...21c5108M,
  2010GApFD.104..207R,2011PhPl...18c2110K,2011PhRvL.106y4502K}.
However, so far, experimental dynamo action driven by a
von K{\'a}rm{\'a}n--like flow is obtained only at the VKS facility and
only when at least one of the flow-driving impellers is made of soft
iron with a large relative permeability.  Kinematic simulations of the
Cadarache dynamo indicate a close linkage between the exclusive 
occurrence of dynamo action in the presence of soft iron impellers and
the observed axisymmetry of the magnetic field
\cite{2010PhRvL.104d4503G,2012NJPh...14e3005G}.  Nevertheless, a fully
satisfactory explanation of the working principle of this dynamo is
still missing and it is still unclear whether the present experiments
will ever be able to achieve growing equatorial dipole modes, which
constitute the magnetic field geometry that has been expected from kinematic
simulations with an axisymmetric flow field.

An improvement of present numerical models may require the explicit
consideration of coherent nonaxisymmetric structures that repeatedly
have been observed in water experiments using a von K\'arm\'an--like
flow driving \cite{2007PhRvL..99e4101D, 2008JFM...601..339R,
  2009PhFl...21b5104C}.  Nonaxisymmetric time-periodic flows with a
dominant azimuthal wave number $m=2$ have also been found in 3D
simulations of {\it{s2t2}} flows in spherical geometry
\cite{2009PhRvE..80e6304R}.  Kinematic dynamo simulations using
various manifestations of these velocity fields showed a surprising
diversity of behavior patterns, however, self-generation of
magnetic energy was found only when the time-dependent flow field was taken into account,
whereas the simulations with the time-averaged flow or with different
snapshots of the velocity field did not exhibit dynamo action.
A similar behavior has also been 
found previously in an ideal two-dimensional model by
the authors of Ref. \cite{dormy_gerard-varet_2008}, who
examined the induction action of a uniform shear flow
perturbed by an periodic variation on intermediate time scales.
The authors found a perpetual amplification even for very small
perturbation amplitudes and concluded that tiny distortions
$\sim \rm{Rm}^{-1}$ can be sufficient to alter the ability of a flow to provide for
dynamo action.   
This type of dynamo action has been attributed to {\it{non-normal growth}} in
Ref. \cite{2008PhRvL.100l8501T}, where it was shown that an appropriate mixing of
nonorthogonal eigenstates through a time-dependent linear operator
can lead to growing modes even if the contributing eigenstates
alone correspond to decaying solutions in a stationary system. 

The main objective of our study is the behavior of the dynamo
efficiency of a cylindrical VKS-like system subject to
nonaxisymmetric velocity perturbations with a single azimuthal
wavenumber $m=2$.  Such velocity modes were observed in a water
experiment conducted at the University of Navarra in Pamplona in order to
analyze the influence of slowly evolving large-scale flow on the
occurrence of dynamo action
\cite{2007PhRvL..99e4101D,ISI_000271475800018}.  Here, we utilize
the essential features of the measured flow field as the basis input
for numerical simulations of the kinematic induction equation.
Typical input parameters that are systematically varied are the flow
amplitude (in terms of the magnetic Reynolds number) and the azimuthal
drift motion of the implied nonaxisymmetric velocity perturbation.
From the simulation data we extract the leading eigenmodes and the
related eigenvalues in terms of growth rates and frequencies that
describe field amplitude modulations and/or azimuthal field drift.
Interestingly, for comparably low drift frequencies of the velocity
perturbation, we, first, observe a phase locking of the magnetic
eigenmode drift with the vortex drift which is replaced, for higher
perturbation drift frequency, by
the appearance of a time-modulated magnetic eigenmode. By analyzing
the involved growth rates and frequencies in the phase-locked regime
and in the modulated regime, we identify the transition between them
as a spectral exceptional point where eigenvalues and eigenfunctions
of two modes coincide \cite{katobook,2004CzJPh..54.1039B}.  The
observed behavior is in close analogy with typical (resonant)
mechanical systems subject to periodic forcing, like, e.g., spinning
disk systems \cite{1992JAM....59..390C}, or to the behavior observed in
the stability study of water waves \cite{1986RSPSA.406..115M}
and we will see, by analyzing the solution of a simple Mathieu
equation, that the observed spectral structure is quite generic
for systems under the influence of periodic forcing.
Comparable effects have also been found in mean-field dynamos of
$\alpha\omega$-type that were designed to explain the bisymmetric
field pattern observed in spiral galaxies.  In these models a periodic
perturbation is caused by density waves due to spiraling arms, and a
parametric resonance (also called {\it{swing excitation}}) is observed
when the frequency of the perturbing velocity pattern is twice the
oscillation frequency of the (axisymmetric) dynamo 
\cite{1990MNRAS.244..714C,1992A&A...264..319S, 1996A&A...308..381M,
  1999A&A...347..860R, 2002MNRAS.334..925S}. 

In the present paper, we show that a facilitation of dynamo action
by periodic flow perturbations is also possible in more complex
three-dimensional models that include magnetic diffusivity and
potentially can be applied to existing dynamo experiments. 
In contrast to the dynamo models from Refs. \cite{2009PhRvE..80e6304R}
and \cite{2008PhRvL.100l8501T} the observed increase of the
growth rate occurs already  
without involving time-periodic states, which makes an interpretation in terms of non-normal
growth (as in Ref. \cite{2008PhRvL.100l8501T}) rather implausible.

\section{Experimental background}
\begin{figure}[h]
\includegraphics[width=14cm]{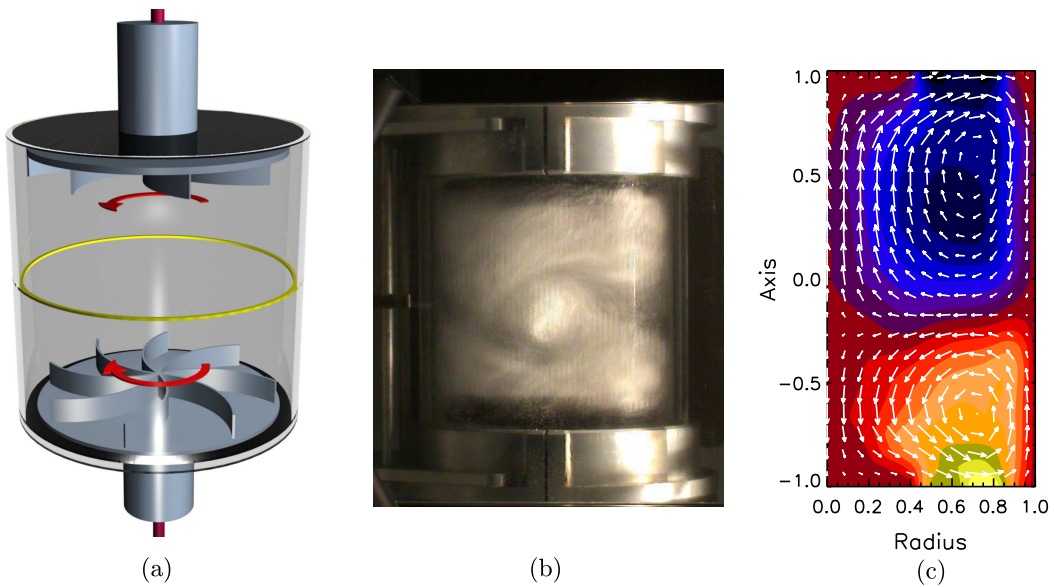}
\caption{\label{fig::exp_flow}(Color) (a) Sketch of the experimental setup. In
  the experiment the von-K\'arm\'an swirling flow is driven by two
  counter-rotating impellers located at the end caps of a cylindrical
  vessel of radius $R= 10{\mbox{ cm}}$ and height $H=20{\mbox{
      cm}}$. The yellow ring denotes the equatorial plane of the
  cylinder.  (b) Instantaneous snapshot of the turbulent fluid
  flow. Note the vortex like structure roughly in the center of the
  vessel.  (c) Mean axisymmetric flow field measured in the
  experiment. The colored structure denotes the toroidal flow
  ($u_{\varphi}$) and the arrows denote the poloidal flow ($u_r,
  u_z$). Although both impellers rotate with the same velocity
  different sizes of the flow cells occur due to spontaneous symmetry
  breaking.}
\end{figure}

The simulations presented below are motivated from a water experiment
that is described in detail in Ref. \cite{2007PhRvL..99e4101D}.  In the
experiment a von K\'arm\'an swirling flow is driven by two
counter-rotating impellers located at the end caps of a cylindrical
vessel of radius $R= 10{\mbox{ cm}}$ and height $H=20{\mbox{ cm}}$
[Fig.~\ref{fig::exp_flow}a].
Both impellers spin with a rotation rate up to $12$ ${\rm{Hz}}$ so
the resulting flow is highly turbulent with flow fluctuations of
the same order as the mean flow [a typical snapshot of the turbulent
flow is shown in Fig.~\ref{fig::exp_flow}(b)].  The velocity field is
measured using {\it{laser Doppler velocimetry}} (LDV) and the mean
velocity field is obtained by averaging the instantaneous flow for
(at least) 100 impeller turns.  The resulting axisymmetric velocity
field consists of two toroidal cells and two meridional recirculating
cells that are roughly restricted to each cylindrical half-space
[Fig.~\ref{fig::exp_flow}(c)].  Comparable structures have been also
observed in water experiments with similar configurations
\cite{2008JFM...601..339R,2009PhFl...21b5104C}.
\begin{figure}[h]
\hspace*{-0.7cm}\includegraphics[width=12cm]{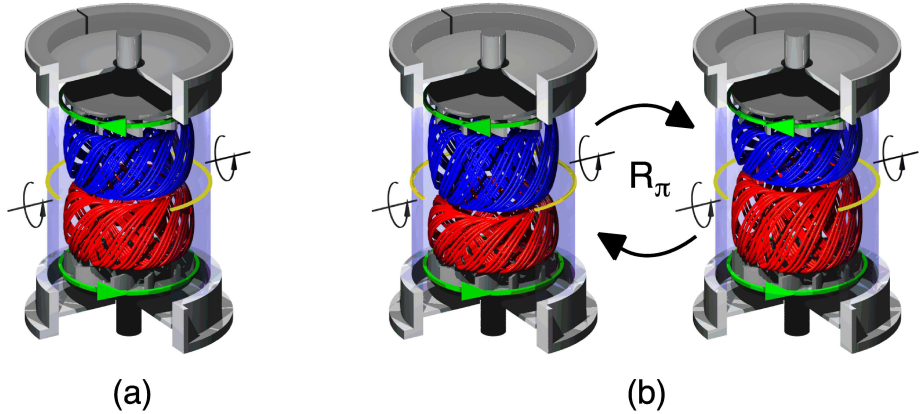}
\caption{(Color online) Reconstruction of averaged trajectories in the experiment.
  The blue (upper)  and red (lower) streamlines represent the averaged trajectories impelled
  by each one of the top and bottom impellers.  (a) Symmetric flow
  configuration.  (b) Equatorial symmetry breaking observed in the
  experiment when ${\rm{Re}} > 10^4$.  Spontaneous jumps between both
  possible states are observed in the experiment.}
\label{fig::symmetries}
\end{figure}
The cylindrical configuration combined with the specific flow driving
imposes a discrete symmetry: The most symmetric flow that can be
obtained between two propellers rotating in opposite directions breaks
the reflection symmetry on any plane that contains the cylinder axis
but preserves the symmetry about a $\pi$ rotation around any diameter
in the equatorial plane [Fig.~\ref{fig::symmetries}(a)].  However, in
the water experiment, this symmetry is broken for ${\rm{Re}} \ga 10^4$
(without internal symmetrizing fixtures) and the measured mean
velocity field becomes asymmetric even when both impellers spin with
equal rotation rates \cite{2007PhRvL..99e4101D}.  As a consequence, one
of the cells [blue (upper cell) or red (lower cell) in Fig.~\ref{fig::symmetries}(b)] becomes
larger than the other.  The averaged velocity field is meta-stable,
and spontaneous jumps of the dominant cell from one side to the other
are observed (so-called {\it{inversions}}) with both possible states
occurring with equal probability.

The transition time for an inversion is $\tau_{\rm{inv}}\approx 10
{\mbox{ s}}$ and the time between consecutive events varies from
minutes to hours following a probability distribution
$\rho(t)=\nicefrac{1}{T_0}e^{-\nicefrac{t}{T_0}}$ with a parameter
$T_0$ determined by the noise intensity. $T_0$ decreases for
increasing Reynolds number, i.e., inversions occur more frequently if
the turbulence is more vigorous.  The inversions of the velocity field
introduce a slow time scale at which the large scale flow and hence
the large scale magnetic field change their structure.  In the long
run, the enhanced dissipation arising from the transition between
different eigenstates (corresponding to the different velocity fields)
can result in a reduced efficiency for the dynamo process when the
mean life time of a velocity state becomes of the same order as the
magnetic diffusion time \cite{2007EPJST.146..313D}.

In addition to breaking of the equatorial symmetry the observed mean
flow also violates the ideal axial symmetry.  Coherent
nonaxisymmetric structures emerge close to the equatorial plane (even
before a turbulent state is reached) and establish a local swirling
flow around an axis perpendicular to the main symmetry axis of the
cylinder.  The nonaxisymmetric flow perturbation is dominated by an
azimuthal wave number $m=2$ and undergoes an azimuthal drift that is
immediately linked to the flow orientation of the dominating toroidal
cell.  The drift frequency of the vortexlike pattern is related to
the maximum azimuthal mean flow velocity $u_{\varphi}^{\rm{max}}$ by
$\omega_{\rm{v}}\approx 0.3 u_{\varphi}^{\rm{max}}/R_{\rm{v}}$, where
$R_{\rm{v}}$ denotes the radius of the maximum vortex velocity ($
R_{\rm{v}}\approx 0.857$ in units of the cylinder radius).  The
mid-size structures provide an additional source of helicity, so 
it is likely that the vortices are involved in the dynamo process when
a conducting fluid like liquid sodium is utilized.

\section{Numerical model}
The temporal development of the magnetic flux density $\vec{B}$
induced from a flow $\vec{u}$ of a conducting fluid is described by
the magnetic induction equation that results from the combination of
Faraday's law, Ohm's law, and Amp{\`e}re's law (without displacement
current):
\begin{equation}
\frac{\partial\vec{B}}{\partial t}=
\nabla\times\left(\vec{u}\times\vec{B}
-\eta\nabla\times\vec{B}\right).
\label{eq::ind}
\end{equation}
In (\ref{eq::ind}) $\eta$ denotes the magnetic diffusivity which is
related to the electrical conductivity $\sigma$ and the vacuum
permeability $\mu_0$ by $\eta=(\mu_0\sigma)^{-1}$.
Equation~(\ref{eq::ind}) is time stepped applying a finite volume
method where a constraint transport scheme ensures the exact treatment
of the solenoidal property of $\vec{B}$ (if the initial field is
divergence free).  Insulating boundary conditions are treated with a
modified boundary integral equation approach which yields the
tangential field components on the boundary from the normal field
components on the whole surface of the computational domain
\cite{2005GApFD..99..481I,2008giesecke_maghyd}.

\subsection{The axisymmetric velocity field}
\begin{figure}[t!]
\includegraphics[width=10cm]{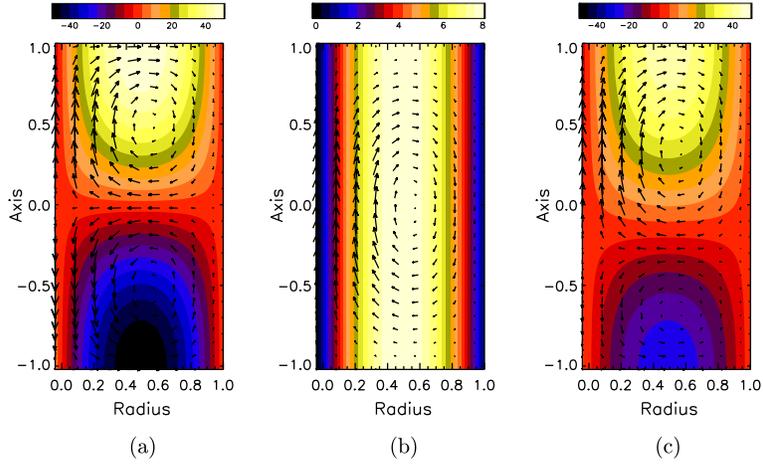}
\caption{\label{fig::flow}(Color online) Prescribed velocity field. From left to
  right: (a) even flow (MND flow) as given by
  Eq.~(\ref{eq::s2t2}), (b) odd flow contribution
  given by equation~(\ref{eq::s1t1}), (c) total flow
  $\vec{u}=\vec{u}^{\rm{e}}+a\vec{u}^{\rm{o}}$ for $a=0.35$. The
  color-coded (gray shaded) pattern represents the azimuthal flow
  ($u_{\varphi}$) and the 
  arrows denote orientation and magnitude of the meridional flow
  ($u_r, u_z$).  }
\end{figure}

We use an analytically prescribed flow field that incorporates the main
characteristics of a von-K\'arm\'an flow and allows a convenient
variation of the equatorial symmetry breaking.  Since the typical
lifetime of the mean flow state is much longer than the time scales
governing the (nonaxisymmetric) flow perturbations the dynamics
introduced by the inversions can be ignored so we assume a
time-independent axisymmetric velocity field.  Equatorial symmetry
breaking is modeled using a basic flow that is composed of two
different axisymmetric parts with different symmetry properties.  The
main part is symmetric with respect to a $\pi$ rotation around any
diameter in the equatorial plane.  This contribution will be called
{\it{even flow}} [Fig.~\ref{fig::flow}(a)]. The geometric structure of
the even flow consists of two counter-oriented toroidal cells and two
recirculating poloidal cells located in each cylindrical half-space.
Mathematically, the even part is prescribed by the so-called MND flow
first proposed by Mari{\'e}, Normand, and Daviaud \cite{2004phfl},
\begin{eqnarray}
u_r^{\rm{e}}(r,z)&=&-\frac{\pi}{H} r(1-r)^2(1+2r)
\cos\!\left({{\frac{2\pi z}{H}}}\right),\nonumber\\ 
u_{\varphi}^{\rm{e}}(r,z)&=&4\epsilon r(1-r)
\sin\left({{\frac{\pi z}{H}}}\right),\label{eq::s2t2}\\ 
u_z^{\rm{e}}(r,z)&=&(1-r)(1+r-5r^2)
\sin\left({{\frac{2\pi z}{H}}}\right),\nonumber
\end{eqnarray}
where $\epsilon$ represents the relation between toroidal and poloidal
flow (here $\epsilon=0.8155$) and $H$ is the total height of the
cylinder (here $H=2.0$).

The second contribution changes its sign when performing the same
$\pi$ transformation and is called {\it{odd flow}}
[Fig.~\ref{fig::flow}(b)].  This contribution is determined by a
single poloidal cell that spreads through the whole cylinder and a
global azimuthal rotation $u_{\varphi}$ (independent of $z$ and
$\varphi$).  The odd flow contribution assumes the same expressions
for $\epsilon$, $H$ and for the radial profile as used for the even
flow~(\ref{eq::s2t2}) and is given by
\begin{eqnarray}
u_r^{\rm{o}}&=&\frac{\pi}{2H} r(1-r)^2(1+2r)
\sin\left(\frac{2\pi z}{H}\right),\nonumber\\ 
u_{\varphi}^{\rm{o}}&=&\epsilon r(1-r),\label{eq::s1t1}\\ 
u_z^{\rm{o}}&=&(1-r)(1+r-5r^2)
\cos^2\left(\frac{\pi z}{H}\right).\nonumber
\end{eqnarray}
The variation of the equatorial symmetry breaking is realized by
multiplying the odd flow contribution~(\ref{eq::s1t1}) with a
weighting factor $a\in [0;1]$ so the total flow field is given by
\begin{equation}
\vec{u}=\vec{u}^e+a\vec{u}^o.\label{eq::total_flow}
\end{equation}
$\vec{u}$ describes a flow field with two different cell sizes with
the even contribution always being dominant [see Fig.~\ref{fig::flow}(c)].
In our model the dominant toroidal eddy is always located close to the
north (top) impeller (at $z=+H/2$) and is characterized by a
counterclockwise oriented azimuthal velocity
(figure~\ref{fig::flow}c).

The combined flow field is scaled allowing a systematically variation
of the (prescribed) magnetic Reynolds number defined as
\begin{equation}
{\rm{Rm}}=\frac{\mathcal{U}\mathcal{R}}{\eta},\label{eq::rm}
\end{equation}
where $\eta$ denotes the magnetic diffusivity, $\mathcal{R}$ a
characteristic length scale (here $\mathcal{R}=1$, the radius of the
cylindrical domain), and $\mathcal{U}$ is the peak velocity defined by
$\mathcal{U}=U^{\max}=\max[(u_r^2+u_{\varphi}^2+u_z^2)^{\nicefrac{1}{2}}]$.

The flow field~(\ref{eq::total_flow}) is qualitatively similar to the
model flow applied in Ref. \cite{2009EL.....8739002G}, where dynamo action
is examined in a VKS-like configuration with impellers rotating at
different speeds.  There are various other possibilities to realize
the equatorial symmetry breaking of the ideal flow field
(\ref{eq::s2t2}) and it turns out that not every property of the
observed flow field can always be reproduced exactly.  For example,
the definition of the flow field~(\ref{eq::total_flow}) does allow an
easy adjustment of the symmetry breaking in terms of the parameter
$a$, but exhibits different azimuthal velocities near the
impellers. In this sense the idealized flow~(\ref{eq::total_flow})
differs from the observed flow, which results from a forcing through
impellers that are in exact counter-rotation, so the flow
velocities close to both impellers are the same.  Such deviations
slightly influence the quantitative outcome of the simulations like,
e.g., the critical magnetic Reynolds number for the onset of dynamo
action; nevertheless, they do not affect the essential conclusions that
will result from our simulations.

\subsection{Modeling of the non-axisymmetric velocity perturbations} 
\begin{figure}[h]
\includegraphics[width=14cm,angle=0]{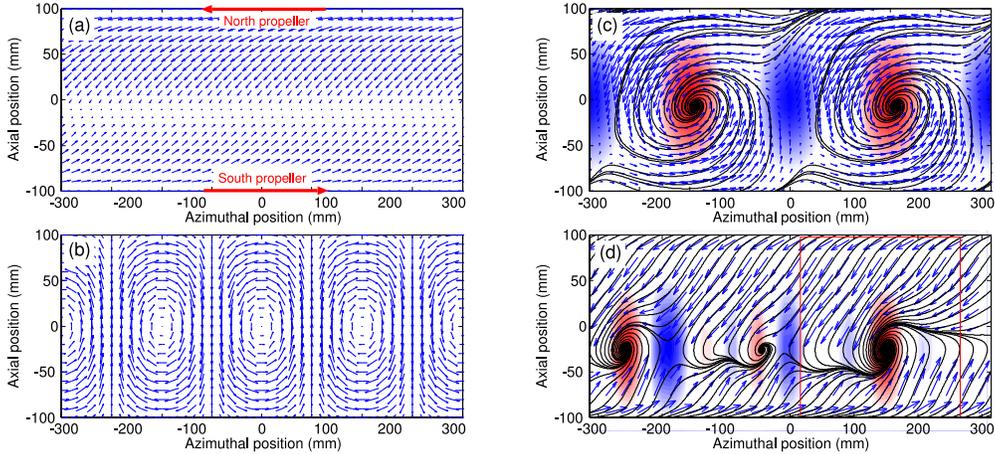}
\caption{(Color online) Mean flow and vortex pattern realized in the model: The
  figures show axisymmetric and nonaxisymmetric contributions of the
  velocity field in the $(\varphi,z)$ plane at $r=0.9$.  Only the
  azimuthal and axial components of the velocity field are presented.
  (a) Odd and even axisymmetric part (Eqs.~(\ref{eq::s2t2})
  and~(\ref{eq::s1t1})).  (b) Vortices according to~(\ref{eq::vortex})
  for $m=2$.  (c) Total velocity field [combining
  Eqs.~(\ref{eq::s2t2}),~(\ref{eq::s1t1}), and~(\ref{eq::vortex})] for 
  $V_m=0.2$. The colored (gray shaded) structures represent the radial component of
  the vorticity of the nonaxisymmetric contribution
  $(\nabla\times\vec{u}^{\rm{v}})_r$ [red (dark gray) respectively blue (light
  gray) corresponds to
  positive repectively negative vorticity].  (d) Same as in (c) for the
  experimental data.}
\label{fig::roll_formation}
\end{figure}

The coherent nonaxisymmetric flow perturbations are located near the
outer wall of the cylinder and close to the equatorial plane where a
strong shear layer emerges that is caused by the opposite azimuthal
flow orientation in each cylindrical half.  The vortices allow a
relaxation of the shear in the equatorial layer in some way similar to
a Kelvin-Helmholtz instability but under turbulent conditions.  The
formation of the vortices in the experiment and the corresponding
implementation in the numerical model are sketched in
Fig.~\ref{fig::roll_formation}.
In the simulations, the vortices are parameterized by analytical
expressions whereby the essential properties -- the azimuthal wave
number, thickness, diameter, and average radial position -- are taken from
the measurements.  The center of the vortices with a typical diameter
of 0.5 (and an aspect ratio of unity) is located around the equatorial
plane with a maximum at a radius of $R_{\rm{v}}=0.857$ (all numbers
are denoted in units of the cylinder radius $R=10\mbox{ }{\rm{cm}}$).
The nonaxisymmetric contribution of the vortices to the velocity
field is explicitly given by
\begin{eqnarray}
u_r^{\rm{v}}&=&0\nonumber\\ 
u_{\varphi}^{\rm{v}}&=&\frac{2V_{\rm{m}}}{m}
\cos(m(\varphi+\omega_{\rm{v}} t))r^2 (1-r)
e^{-{[(r-r_0)/\sigma]^2}}\sin(4z)\label{eq::vortex}\\ 
u_z^{\rm{v}}&=&-2V_{\rm{m}}
\sin(m(\varphi+\omega_{\rm{v}} t))r^2 (1-r) e^{-{[(r-r_0)/\sigma]^2}}
\cos^2(2z),\nonumber
\end{eqnarray}
with the axial coordinate $z$ restricted to the interval $z\in
[-\pi/4;\pi/4]$. In~(\ref{eq::vortex}) $V_{\rm{m}}$ denotes the
magnitude of the nonaxisymmetric perturbation and the quantities
$\sigma$ and $r_0$ are estimated from the experimental observations
(here always $r_0=0.9$ and $\sigma=0.12$).  The azimuthal drift
frequency of the vortex perturbation is denoted by $\omega_{\rm{v}}$,
which in the experiment is linked to the azimuthal velocity of the
dominant mean flow cell.  In the system of units applied in the
simulations the observed vortex drift frequency $\sim 0.3
u_{\varphi}^{\rm{max}}/R_{\rm{v}}$ corresponds to
$\omega_{\rm{v}}\approx 17.1\, (14.9, 14.0, 13.1)$ for $a=0\,(0.62,
0.84, 1.00)$.  Note that the drift frequency of the nonaxisymmetric
{\it{velocity pattern}} with an azimuthal wave number $m$ is related
to the previously defined vortex frequency by
$\Omega_{\rm{p}}=m\omega_{\rm{v}}$.  In the water experiment the
vortices exhibit an intermittent behavior, e.g., sudden jumps to a
state with a different wave number ($m=3$ or $m=4$) or fluctuations of
the vortex drift. Furthermore, the locations of the vortices undergo
variations that are connected to the largest fluctuations of the mean
velocity field. Those fluctuations appear centered around the shear
layers and close to the wall.  These temporal alterations are not
considered in our numerical model, where we restrict our examinations
to the case $m=2$ which corresponds to the most probable configuration
observed in the experiment.  In all cases the disturbance introduced
by the non-axisymmetric component remains only weak and has no
influence on the actual $\rm{Rm}$ given by the maximum of the modulus
of $\vec{u}$.

\section{Results}
In the kinematic approach the back-reaction of the magnetic field on
the flow is neglected so the solution of the induction Eq.
(\ref{eq::ind}) represents a linear problem, which in principle could
be solved with the ansatz 
\begin{equation}
\vec{B}(\vec{r}, t) =\vec{B}_0(\vec{r})e^{\lambda t}.\label{eq::exp_time}
\end{equation}
A dynamo solution is obtained if
the magnetic energy density $E_{\rm{mag}}=(2\mu_0)^{-1}|\vec{B}|^2$
grows exponentially, $\propto e^{2\lambda t}$.  In general, $\lambda$ is
a complex quantity, $\lambda=\gamma+i\omega$, where $\gamma$ denotes the
field amplitude growth rate and $\omega$ denotes an oscillation or
drift frequency.  
In case of a time-dependent velocity perturbation with a (given) period
$T$, the general form of the solution differs from~(\ref{eq::exp_time})
and the proper time dependence follows from Floquet
theory,
\begin{equation}
\vec{B}(\vec{r}, t)\sim \sum_i e^{\mu_i t}\vec{P}_i(\vec{r},t),
\end{equation}
where $\vec{P}_i(\vec{r},t)=\vec{P}_i(\vec{r},t+T)$ has period $T$ and
$\rho_i=e^{\mu_i t}$ are the so-called {\it{Floquet multipliers}}
(e.g. Refs. \cite{book_joseph,book_guck}).
In the following, the growth rates represent magnetic field amplitude growth
rates for the $(m=1)$ mode and the applied time scale is given by the
magnetic diffusion time $\tau_{\eta}=R^2/\eta$.
In accordance with Cowling's anti dynamo theorem,
dynamo solutions generated by a prescribed mean axisymmetric flow
necessarily yield a nonaxisymmetric field.  In all simulations
presented below the ($m=1$) mode is the dominant field contribution so
the magnetic eigenmode behaves $\propto\cos\varphi$ and
represents a dipole oriented along the equatorial plane.

\subsection{Equatorial symmetry breaking}
\begin{figure}[b!]
\includegraphics[width=6cm]{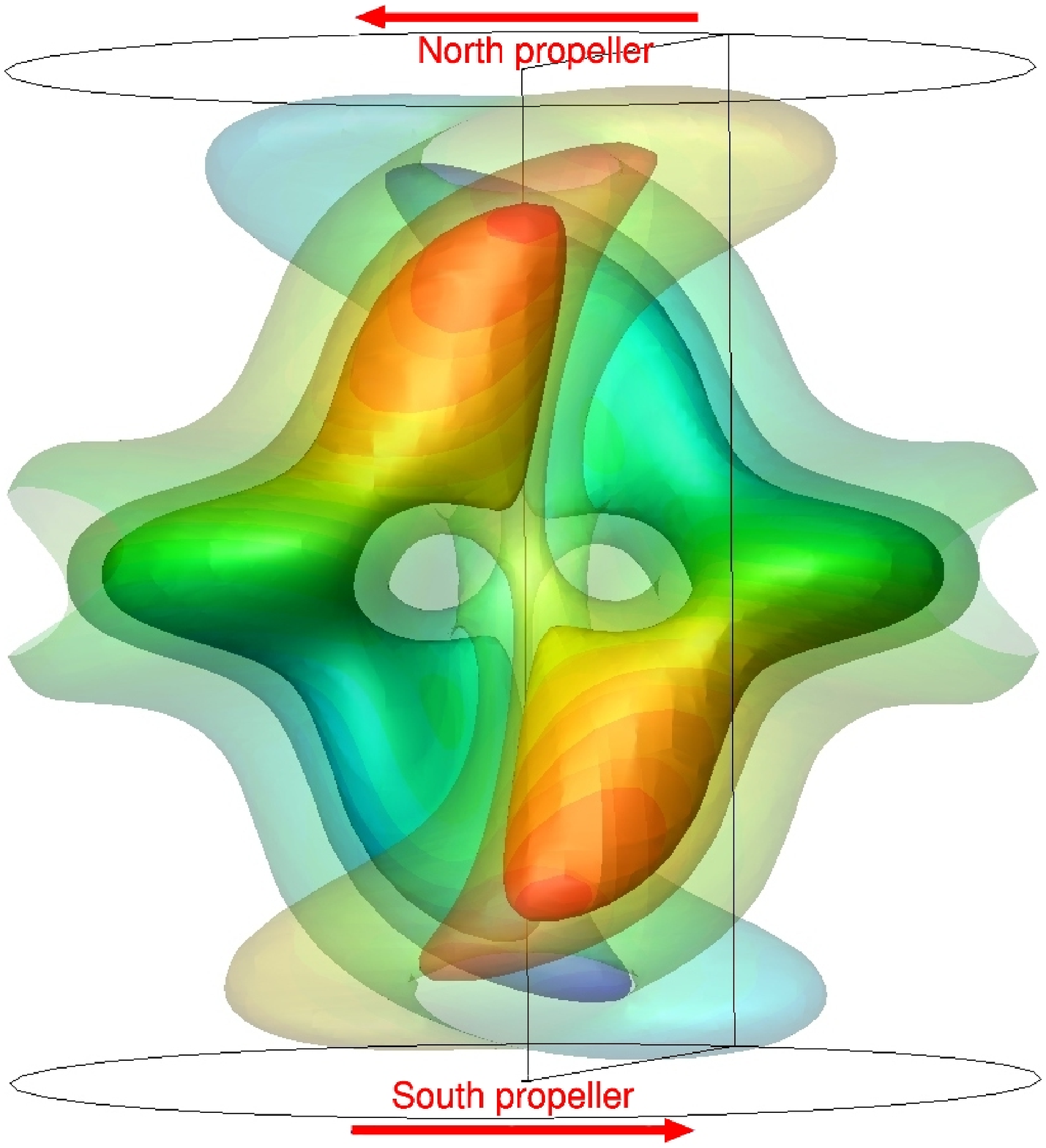}
\hspace*{1cm}
\includegraphics[width=6cm]{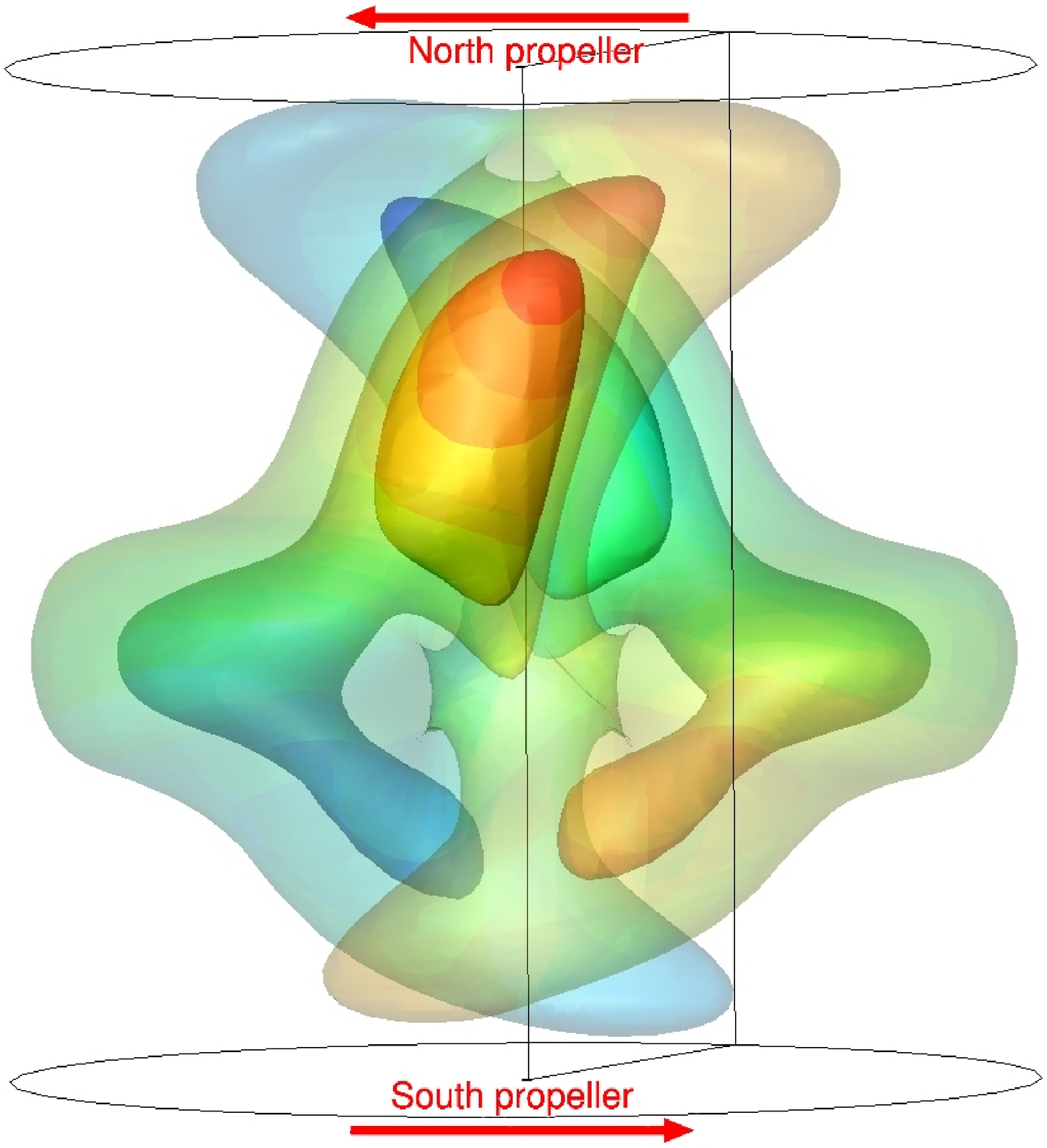}
\caption{(Color) Geometric structure of the dynamo eigenmode.  Left: No
  equatorial symmetry breaking ($a=0$), ${\rm{Rm}}=70$.  Right:
  $a=0.62, {\rm{Rm}}=70$.  The isosurfaces show the magnetic energy
  density at 10\%, 30\%, and 50\% of the respective maximum value.  The
  colored contours denote the azimuthal magnetic field $B_{\varphi}$.
  Note the magnetic energy concentration in the upper part of the
  cylinder, which contains the dominant flow cell in case of equatorial
  symmetry breaking (right panel).}\label{fig::field_pattern}
\end{figure}

\begin{figure*}[h!]
\includegraphics[width=8.00cm]{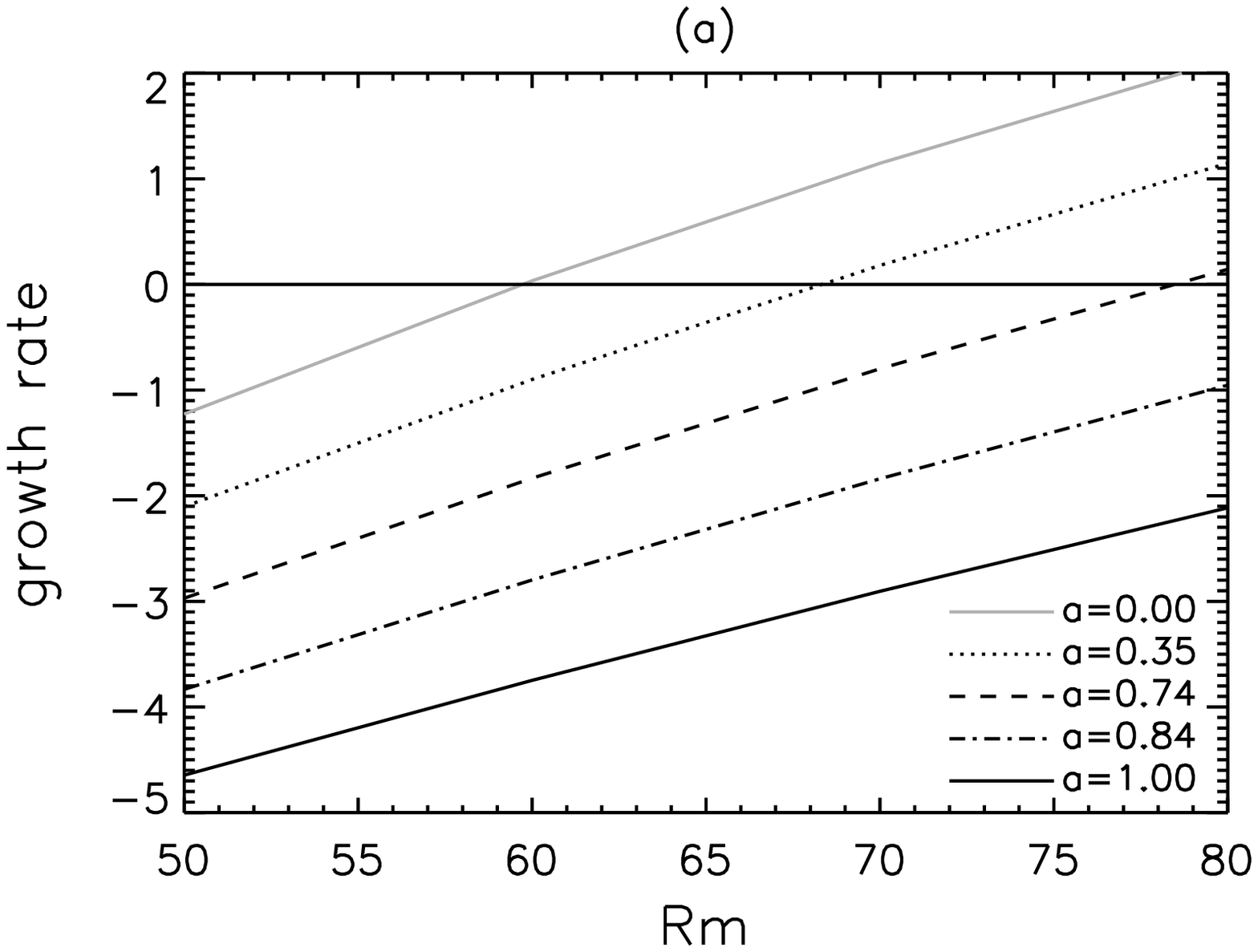}
\includegraphics[width=8.00cm]{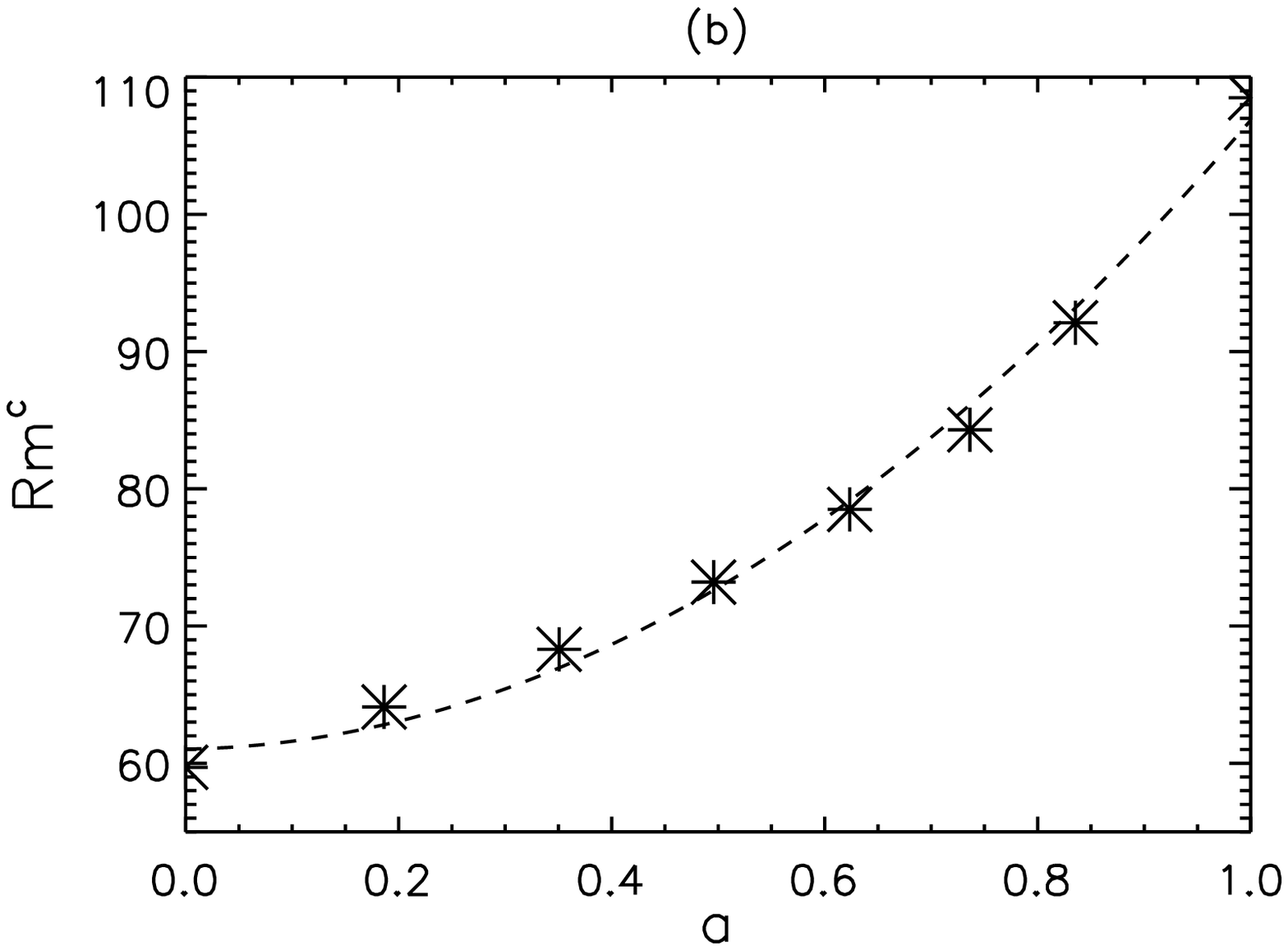}
\caption{ (a) Growth rate versus ${\rm{Rm}}$ for various equatorial
  symmetry breaking $a$; (b) critical magnetic Reynolds number versus
  $a$. The dashed curve denotes a fit by a polynomial of degree 2.
}\label{fig::gr_drift_vs_rm}
\end{figure*}

For a sufficiently large magnetic Reynolds number the flow
field~(\ref{eq::total_flow}) drives a dynamo with a typical structure
shown in Fig.~\ref{fig::field_pattern}, which displays the eigenmodes
for the undisturbed flow field ($a=0$) and the model case $a=0.62$.
In both cases, the geometry is dominated by two interleaved
banana-cell-like structures.  However, in case of equatorial symmetry breaking ($a
= 0.62$, Fig. 5, right panel) a slight concentration of magnetic energy is
observed in the upper cylindrical half-space containing the dominant flow
cell.  The breaking of the ideal equatorial symmetry suppresses dynamo
action: When a symmetry breaking flow contribution is added, the
critical magnetic Reynolds number for the onset of dynamo action
roughly increases $\propto a^2$ from ${\rm{Rm}}^{\rm{c}}\approx 50.5$
for $a=0$ to ${\rm{Rm}}^{\rm{c}}=109.5$ at $a=1$
[Figs.~\ref{fig::gr_drift_vs_rm}(a) \& 6(b)].

\begin{figure*}[h!]
\includegraphics[width=8.00cm]{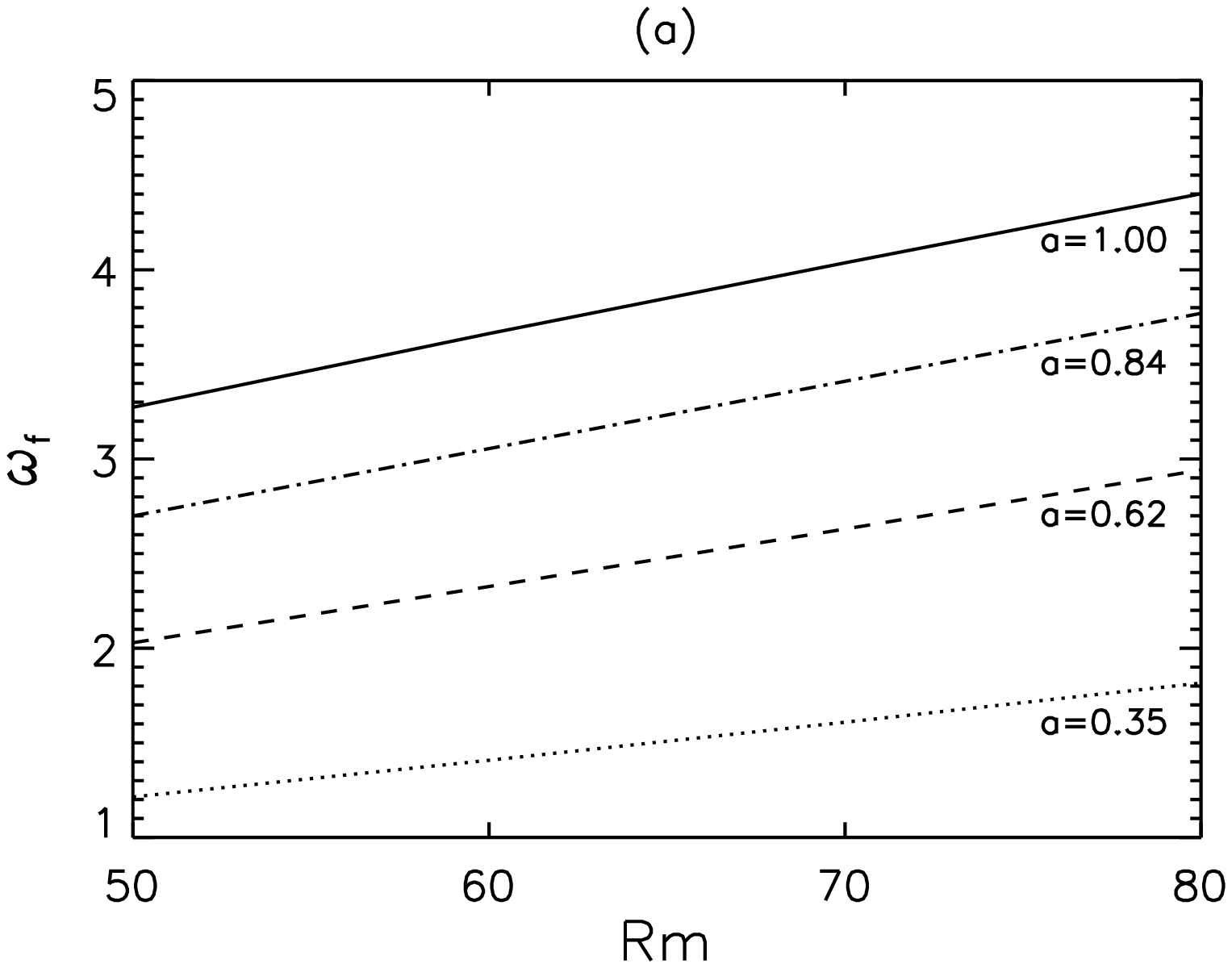}
\includegraphics[width=8.00cm]{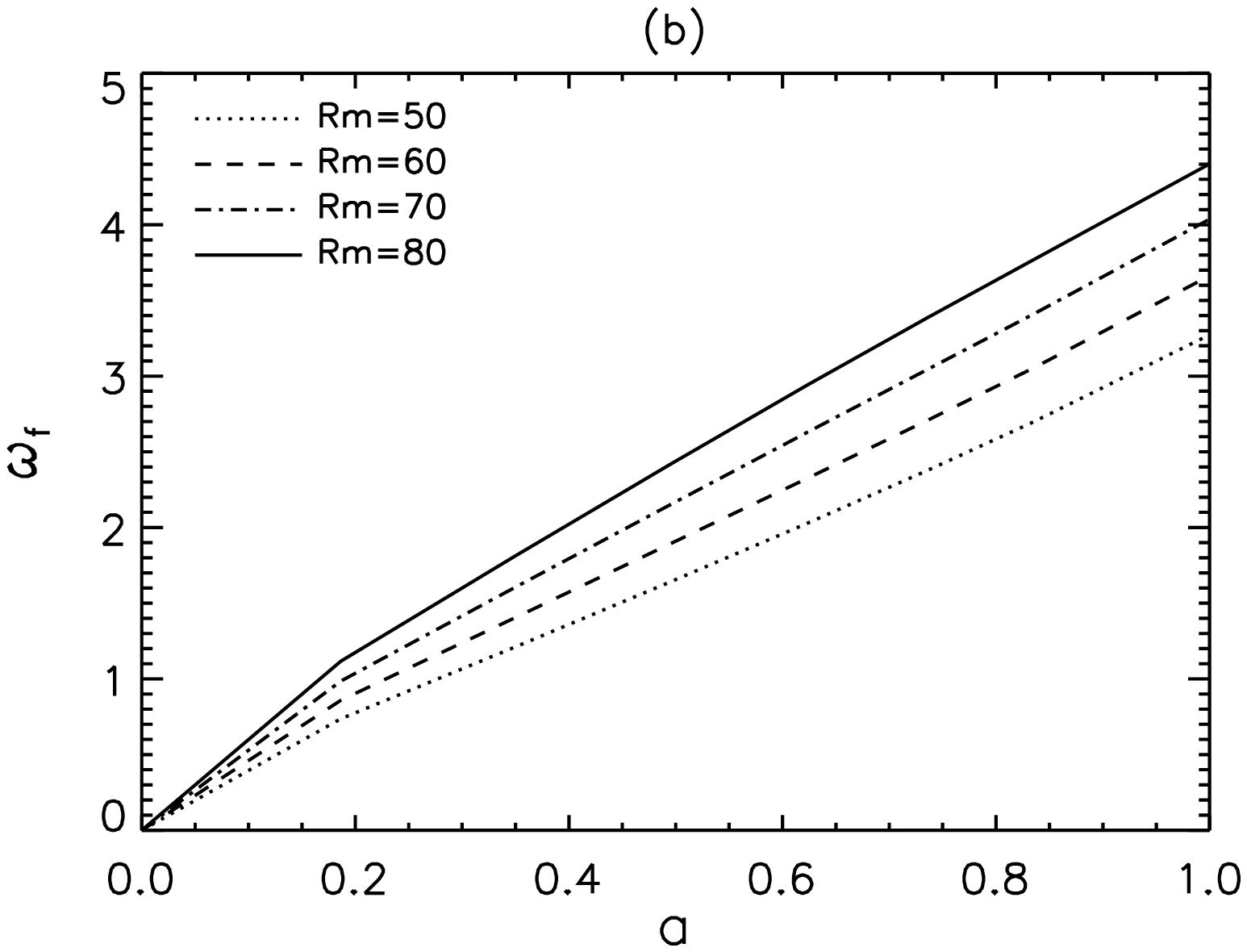}
\caption{ (a) Drift frequency of the $(m=1)$ mode versus ${\rm{Rm}}$
  for various $a$.  (b) Drift frequency of the $(m=1)$ mode versus
  equatorial symmetry breaking $a$. Note the slight deviation from the
  linear behavior for small $a$.  }\label{fig::freq_vs_rm}
\end{figure*}

In the ideal symmetric case ($a=0$), the eigenmode shows no time
dependence except the exponential decay.  With symmetry breaking, the
$(m=1$) mode exhibits an azimuthal drift around the cylinder axis,
i.e.  $\vec{B} \sim\vec{B}_0(r,z)e^{\lambda
  t}\cos(\varphi-\omega_{\rm{f}}t)$ with the magnetic field drift
frequency $\omega_{\rm{f}}$ increasing nearly linearly with the degree
of asymmetry $a$ and with ${\rm{Rm}}$ [Figs.~\ref{fig::freq_vs_rm}(a)
and (b)]. $\omega_{\rm{f}}$ is constant in time and is oriented opposite
to the azimuthal flow in the dominant cell.  For axisymmetric velocity
fields [as $\vec{u}^{\rm{e}}$ and $\vec{u}^{\rm{o}}$,
Eqs.~\ref{eq::s2t2} and \ref{eq::s1t1}) poloidal and toroidal velocity
components are decoupled, and their shear layer may have independent
locations. The relative position of these shear layers controls the
magnitude (and orientation) of the magnetic field drift.  For example,
an exclusive increment of the odd azimuthal flow contribution (which
only shifts the toroidal shear layer) would alter the drift
frequency. The change might be quite dramatic and even a reversal of
the drift orientation is possible with the eigenmode corotating with
the dominant azimuthal flow, as observed, for example, in
Ref. \cite{2003EPJB...33..469M}.  A systematic study of the influence of
this effect is under way.
\subsection{Impact of non-axisymmetric velocity perturbations}
The presence of non-axisymmetric velocity components results in a
coupling between different azimuthal magnetic field modes that are
separated in the purely axisymmetric problem.  Hence the
characterization of the dynamo eigenmode with a single azimuthal
wavenumber is no longer correct because (in principle) all azimuthal
wave numbers are linked.  In the following, we examine the impact of a
non-axisymmetric flow perturbation with a (single) wave number $m=2$
so two classes of magnetic eigenmodes arise which incorporate
even azimuthal wave numbers ($m=0, 2, 4,...$) or odd azimuthal wave
numbers ($m=1,3,5,...$), respectively.  In our model, only the second
class with odd wave numbers is relevant, whereas the even modes
are not relevant for our problem because they decay on a faster
time scale.

In the following, we keep the amplitude of the nonaxisymmetric
velocity contribution fixed at $V_{\rm{m}} =0.3U^{\rm{max}}$ which
roughly corresponds to the value observed in the experiments.  In this
regime the presence of the non-axisymmetric velocity component neither
changes the global (axisymmetric) flow topology nor the actual
magnetic Reynolds number.  We further assume that the drift frequency
of the vortex, $\omega_{\rm{v}}$, is a free parameter that is
systematically varied in the interval $\omega_{\rm{v}}\in [-60,+60]$.
The maximum or minimum values correspond to a vortex drift
{\it{velocity}} approximately equal to the maximum azimuthal flow
velocity (recall that in the water experiment the observed vortex
drift velocity is roughly $0.3u_{\varphi}^{\max}$).  We also examined
the artificial case with a vortex drift orientation opposite to the
azimuthal flow of the dominant cell. These cases are denoted by
negative frequencies.
\begin{figure}[h]
\includegraphics[width=8cm]{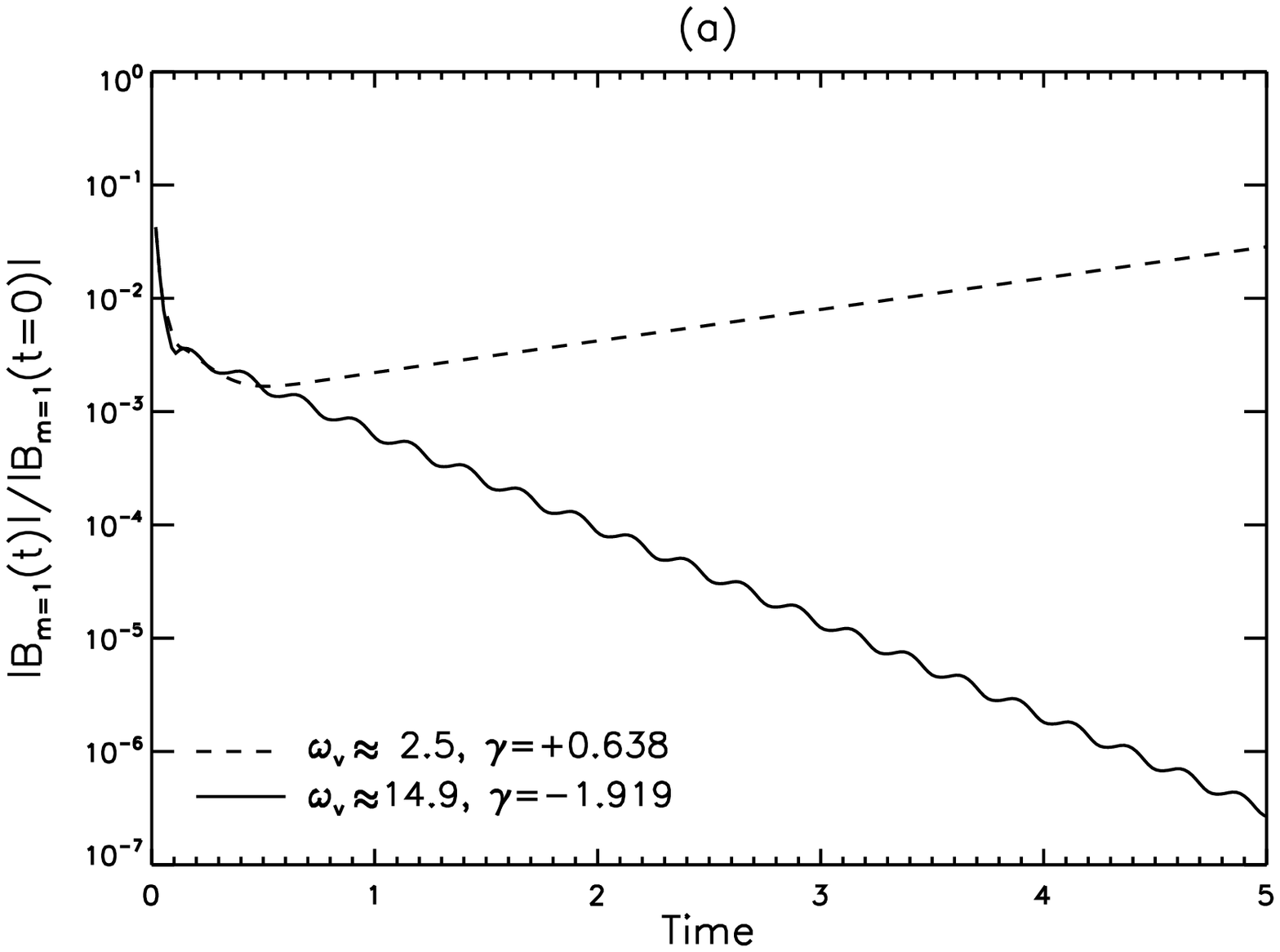}
\includegraphics[width=8cm]{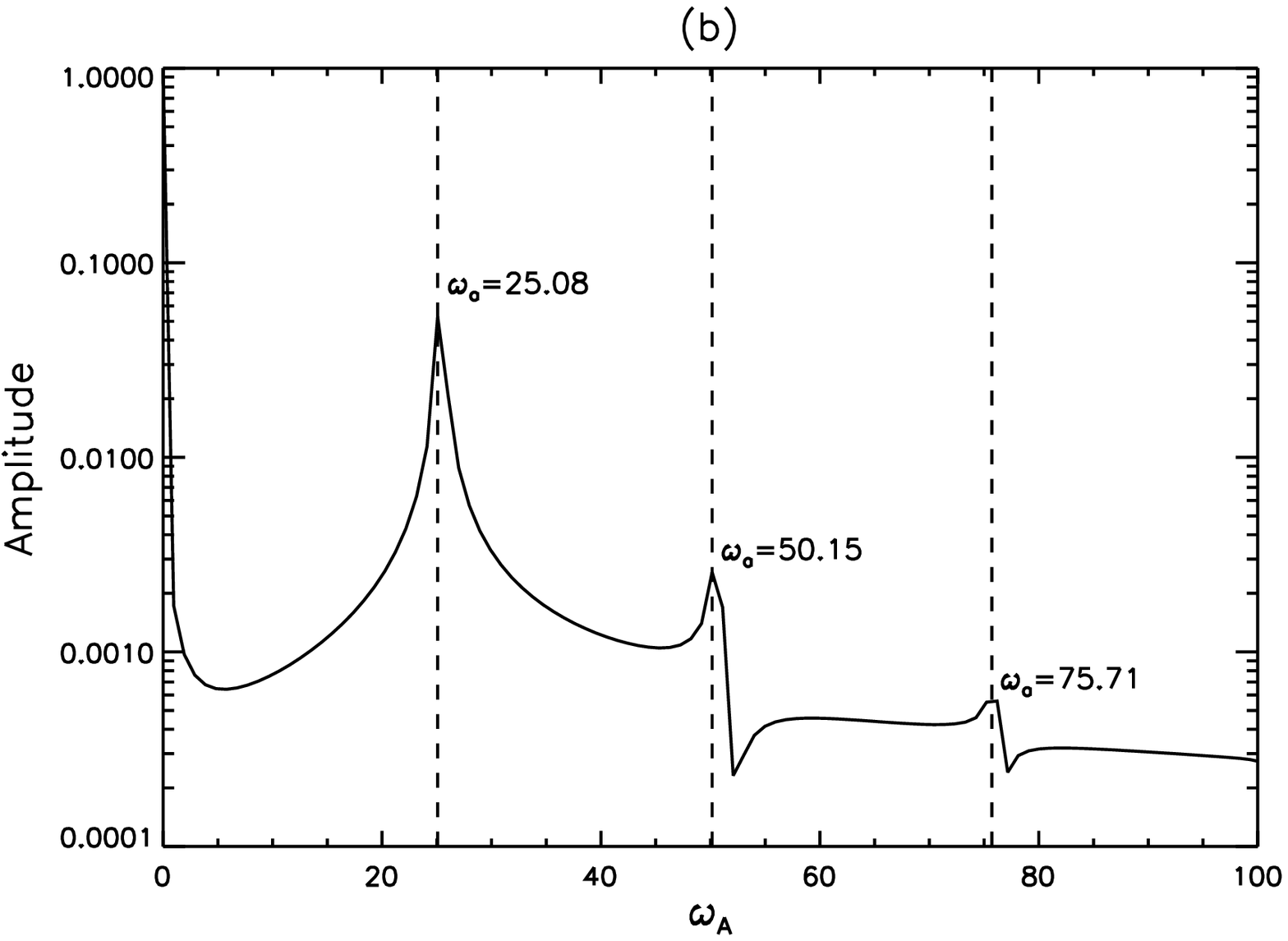}
\caption{ (a) Time dependence of the magnetic field amplitude for
  ${\rm{Rm}}=70$ and $a=0.62$. (b) Fourier spectrum for the case
  $\omega_{\rm{v}}=14.9$ (after eliminating the exponential
  decay). The various peaks denote the modulation frequency of the
  field amplitude $\omega_{\rm{a}}\approx 25.08$ (and its overtones).
\label{fig::gr_vs_time}}
\end{figure}
The temporal development of the amplitude of the dominant eigenmode
for two different vortex drift frequencies, but similar ${\rm{Rm}}$,
is shown in Fig.~\ref{fig::gr_vs_time} (dashed curve:
$\omega_{\rm{v}}\approx 2.5$; solid curve: $\omega_{\rm{v}}\approx
14.9$).  Both curves show completely different behavior: For
$\omega_{\rm{v}}\approx 14.9$ the solution decays and its amplitude is
modulated with a frequency of $\omega_{\rm{a}}\approx 25.08$ (solid
curve), whereas at $\omega_{\rm{v}}\approx 2.5$ the field amplitude
does not show any time dependence except the exponential growth
(dashed curve).  Regardless of the distinct temporal behavior for both
cases the field structure is quite similar (except the pulsating
character in the modulated case) and remains close to the pattern
already observed in the previously discussed unperturbed configuration
(Fig.~\ref{fig::oscmode}).
\begin{figure}[b!]
\hspace*{-1cm}
\begin{minipage}{4.5cm}
\includegraphics[width=4.2cm]{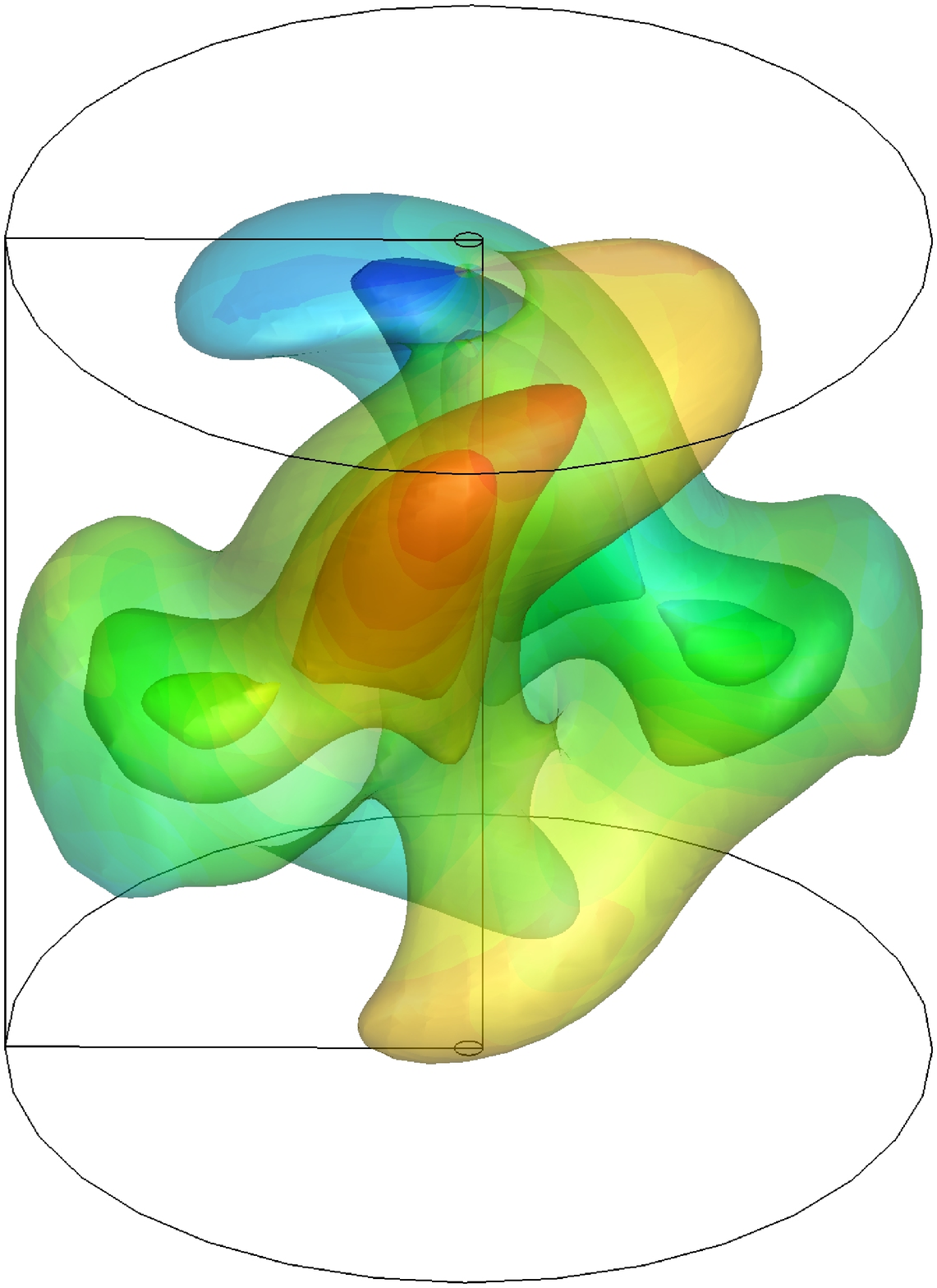}
\end{minipage}
\hspace*{1.5cm}
\begin{minipage}{6cm}
\includegraphics[width=6cm,angle=-90]{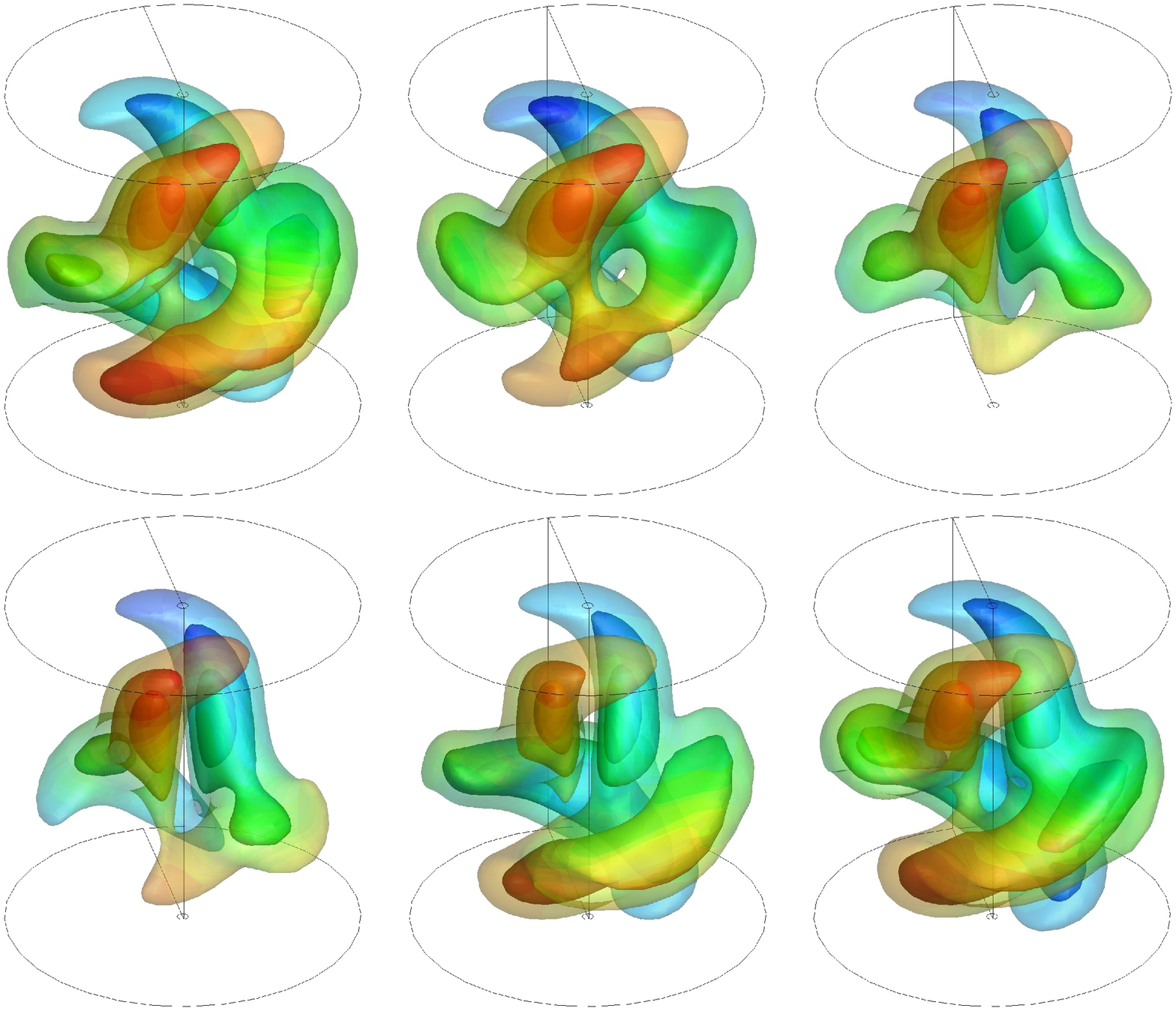}
\end{minipage}
\caption{(Color online) Left: Field structure in the resonant regime (${\rm{Rm}}=70,
  a=0.62, \omega_{\rm{v}}=2.5$, corresponding to the dashed curve in
  the left panel of Fig.~\ref{fig::gr_vs_time}).  Right: Time series
  of a typical solution with amplitude modulation (${\rm{Rm}}=70$,
  $a=0.62$, $\omega_{\rm{v}}\approx 14.9$, corresponding to the solid
  curve in the left panel of Fig.~\ref{fig::gr_vs_time}).  The
  isosurfaces show the magnetic energy density at 10\%, 30\%, and 50\%
  of the respective maximum value and the colored (gray shaded) contours denote the
  azimuthal magnetic field $B_{\varphi}$.  The time series covers one
  modulation period with each snapshot scaled by the respective
  maximum so the effects of exponential decay or growth and the
  amplitude modulations are eliminated.  The period of the amplitude
  modulation is much shorter than the time scale of the magnetic field
  drift so the phase of the eigenmode (i.e., its orientation in
  space) remains nearly constant.  }
\label{fig::oscmode}
\end{figure}

\begin{figure}[t!]
\includegraphics[width=8cm]{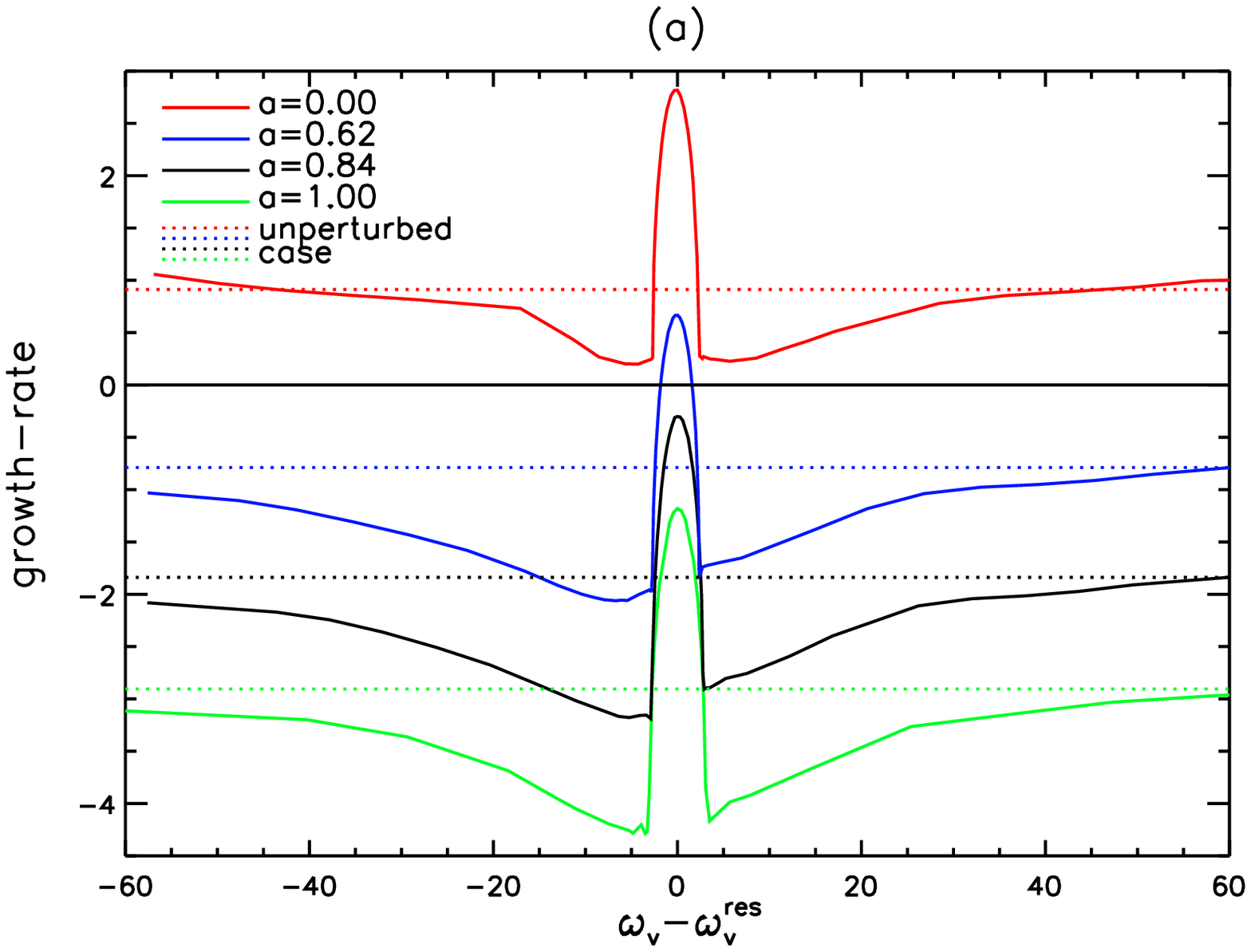}
\includegraphics[width=8cm]{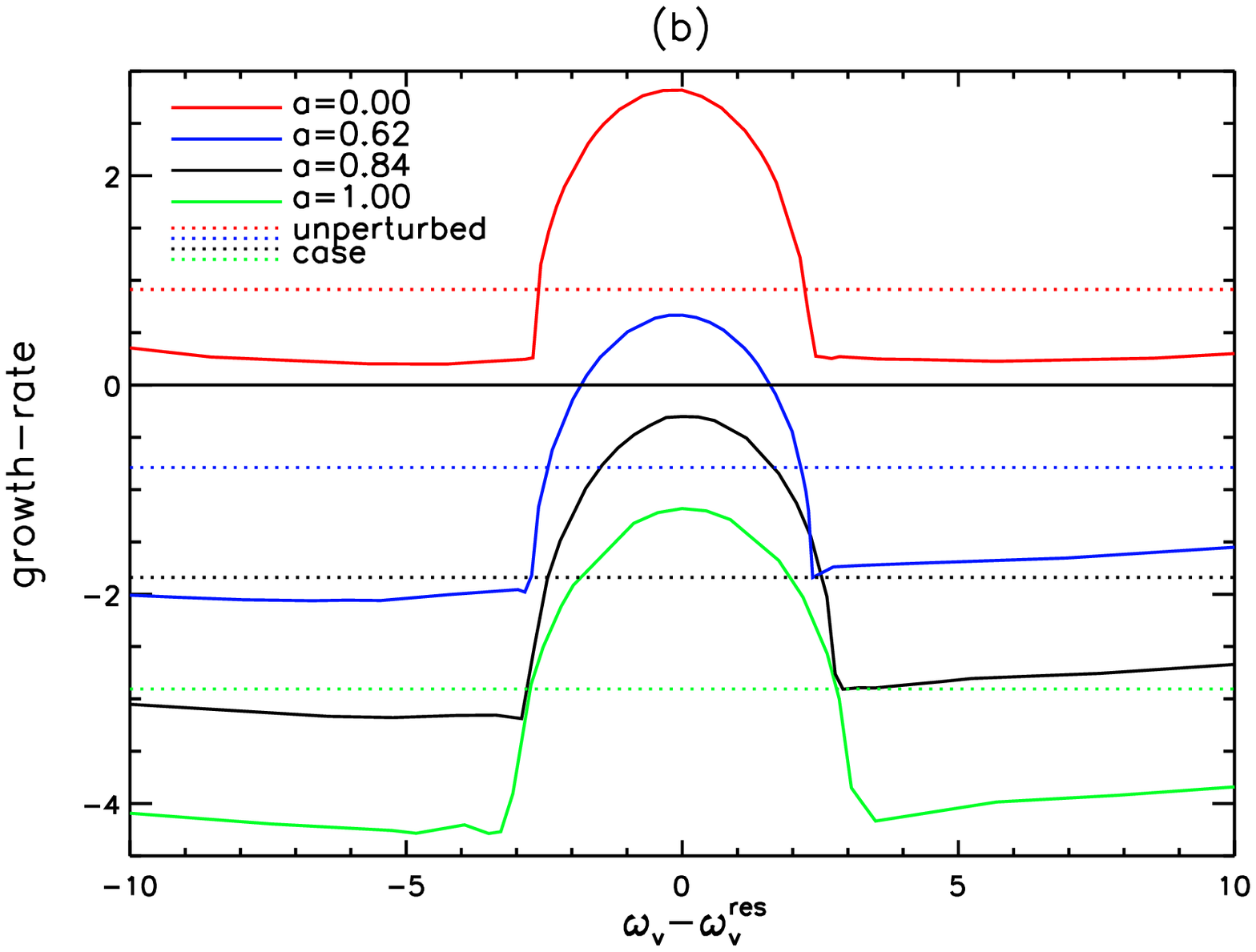}
\caption{\label{fig::gr_vs_omegavortex}(Color online) (a) Growth rate versus vortex
  drift frequency for ${\rm{Rm}}=70$ and $V_{\rm{m}}=0.3U^{\rm{max}}$.
  In the abscissa the resonance frequency $\omega^{\rm{res}}_{\rm{v}}$
  has been subtracted so all curves are centered at the origin.
  The horizontal dotted lines denote the growth rates without
  non-axisymmetric perturbation.  (b) Enlarged section of the resonant
  regime. The growth-rates in the resonant regime correspond to
  nonoscillating $m=1$ eigenmodes, whereas outside of this window the
  solutions exhibit a modulation of the field amplitude [see also
  Fig.~\ref{fig::freq_vs_avgdrift}(c)].  }
\end{figure}

The occurrence of nonaxisymmetric solutions with oscillating energy
is a clear indication for the presence of two distinct azimuthal modes
with the same growth rate (i.e., real part of the eigenvalues) but with
different frequencies (imaginary part of eigenvalues).  In fact,
simulations that cover a sufficient number of periods exhibit
characteristic beat patterns, indicating that the frequencies of the
superimposed modes are very close.  We observe field amplitude
modulations independently of the degree of equatorial symmetry
breaking for a broad range of vortex drift frequencies, whereas
unmodulated solutions only occur within a narrow window of relatively
slow vortex drift frequencies.  The unmodulated solutions are further
characterized by a sharp maximum for the corresponding growth rates
(Fig.~\ref{fig::gr_vs_omegavortex}), typical for parametric
resonance.
Characteristic properties of the resonant regime and a comparison to
values in the unperturbed state are denoted in Table~\ref{tab::freq}.
\begin{table}[h!]
\caption{Characteristics properties of the resonant window for
  different values of the equatorial symmetry breaking $a$.  All data
  stems from runs with ${\rm{Rm}}=70$. The critical magnetic Reynolds
  number (column 6 \& 7) has been estimated from inter/extrapolation
  of simulation runs with ${\rm{Rm}}=50, 70$ and
  $90$.\label{tab::freq}}
\begin{tabular}{c|c|c|c|c|c|c|c|c}
$a$ & {\begin{minipage}{1.8cm}{maximum\\growth rate}\end{minipage}} 
    & {\begin{minipage}{1.8cm}{$\omega^{\rm{res}}_{\rm{v}}$}\end{minipage}}
    & {\begin{minipage}{1.8cm}{unperturbed\\growth rate}\end{minipage}}
    & {\begin{minipage}{1.8cm}{$\omega_{\rm{f}}^{0}$}\end{minipage}} 
    & {\begin{minipage}{1.2cm}{${\rm{Rm}}^{\rm{c}}$\\without vortex}\end{minipage}} 
    & {\begin{minipage}{1.2cm}{${\rm{Rm}}^{\rm{c}}$ at resonance}\end{minipage}} 
    & {\begin{minipage}{1.8cm}{relative\\reduction\\of ${\rm{Rm}}^{\rm{c}}$}\end{minipage}}
    & {\begin{minipage}{2.2cm}{width of\\unmodulated regime $\Delta\omega_{\rm{v}}$}\end{minipage}} 
\\[0.5cm] 
\hline 
0.00 & $+2.817$ & $-0.00$ & $+0.912$ & $\quad 0.00$ & 59.7 & 50.5 & 15.4\% & 5.06\\ 
0.62 & $+0.667$ & $-1.99$ & $-0.798$ & $-2.63$ & 78.5 & 64.8 & 17.5\% & 5.11\\ 
0.84 & $-0.302$ & $-2.91$ & $-1.839$ & $-3.41$ & 92.1 & 72.9 & 20.8\% & 5.63\\ 
1.00 & $-1.181$ & $-3.50$ & $-2.905$ & $-4.04$ & 108.5 & 82.7 & 23.8\%& 6.12\\
\end{tabular}
\end{table}
The location of the resonance maximum (denoted by
$\omega_{\rm{v}}^{\rm{res}}$) is roughly determined by the field drift
frequency of the magnetic eigenmode in the unperturbed case (from now
labeled with $\omega_{\rm{f}}^0$).  However, a constant gap is
observed in runs with equatorial symmetry breaking ($a>0$) resulting
in $\omega_{\rm{v}}^{\rm{res}}\approx (\omega_{\rm{f}}^0+0.5)$,
whereas the case $a=0$ remains a particular case with
$\omega^{\rm{res}}_{\rm{v}}=0$.  The shift of the resonance frequency
with respect to the fundamental frequency in the unperturbed problem
probably follows from the interaction of vortex structure and
{\it{drifting}} eigenmode similar to the shift of the resonance
frequency in periodically forced mechanical systems with damping.
Outside of the resonance, the growth-rate is suppressed compared to
the unperturbed case (denoted by the dotted horizontal line in
Fig.~\ref{fig::gr_vs_omegavortex}).  Only for very large drift
velocities is the value of the purely axisymmetric case obtained
again, i.e., the impact of the vortices vanishes when their drift
frequency becomes too large.  The width of the resonant regime
slightly increases with the symmetry-breaking parameter $a$ and so
does the relative enhancement of the growth rates and, accordingly, the
relative reduction of the critical ${\rm{Rm}}$.

\begin{figure}[h!]
\includegraphics[width=8cm]{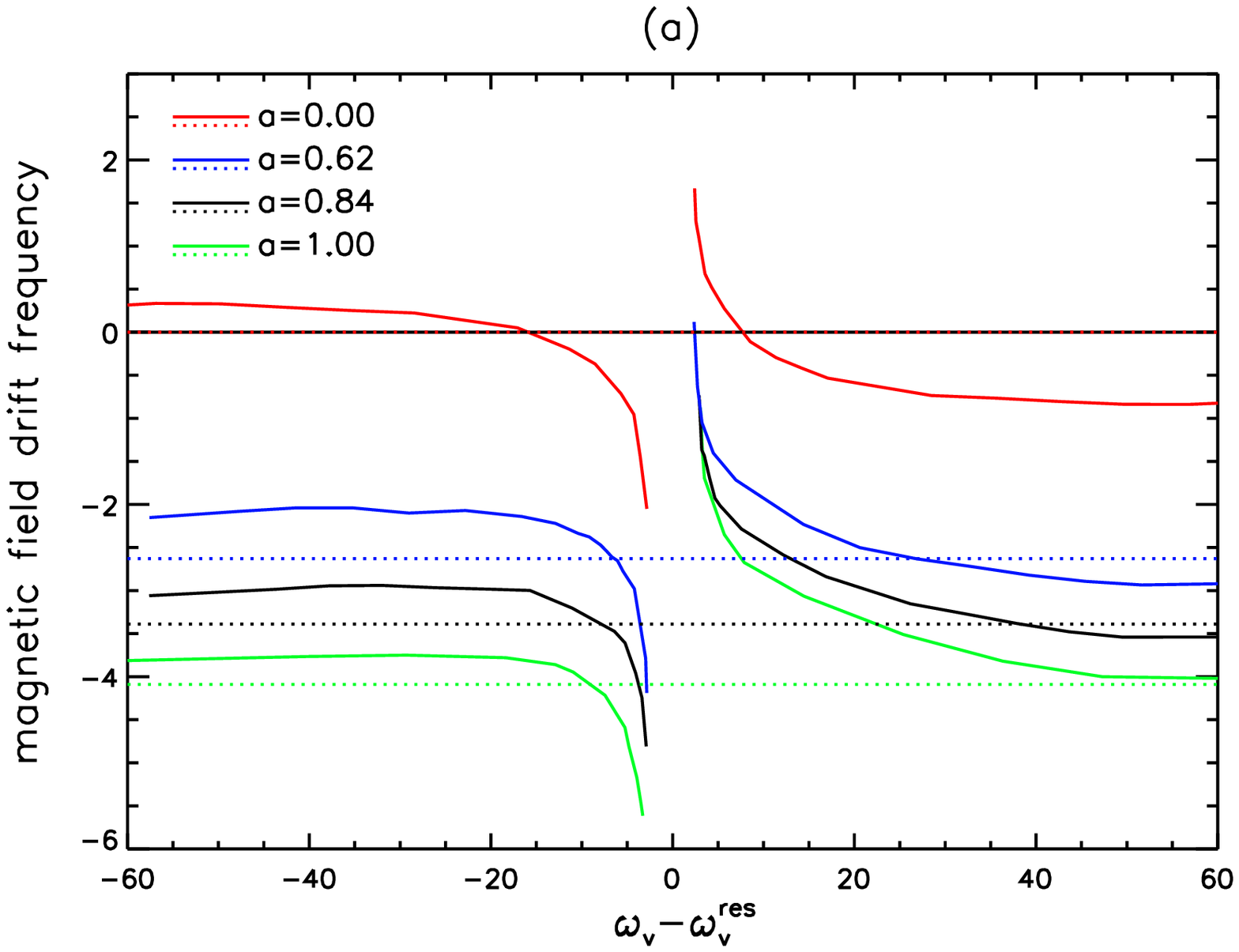} \\[0.3cm]
\includegraphics[width=8cm]{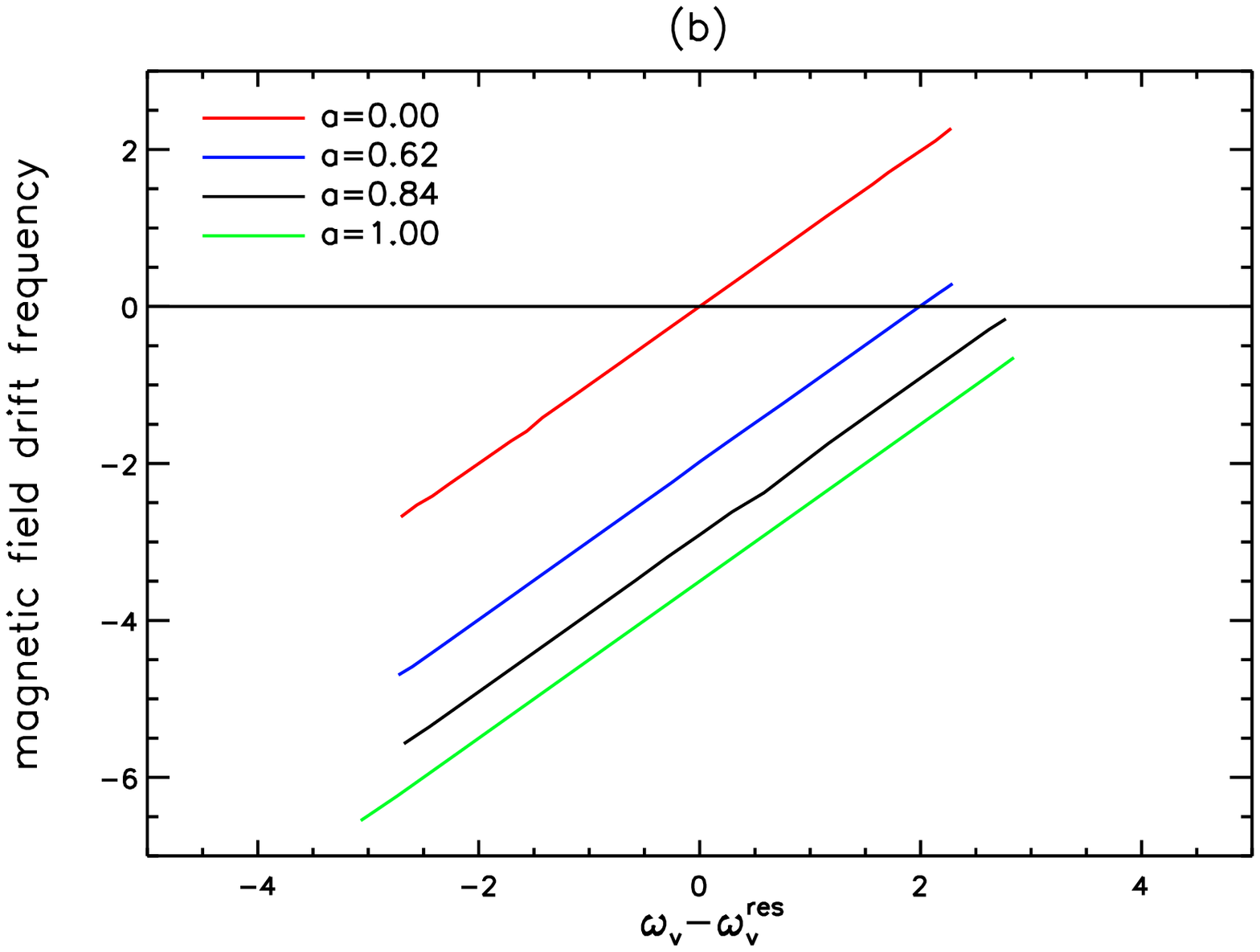} \\[0.3cm]
\includegraphics[width=8cm]{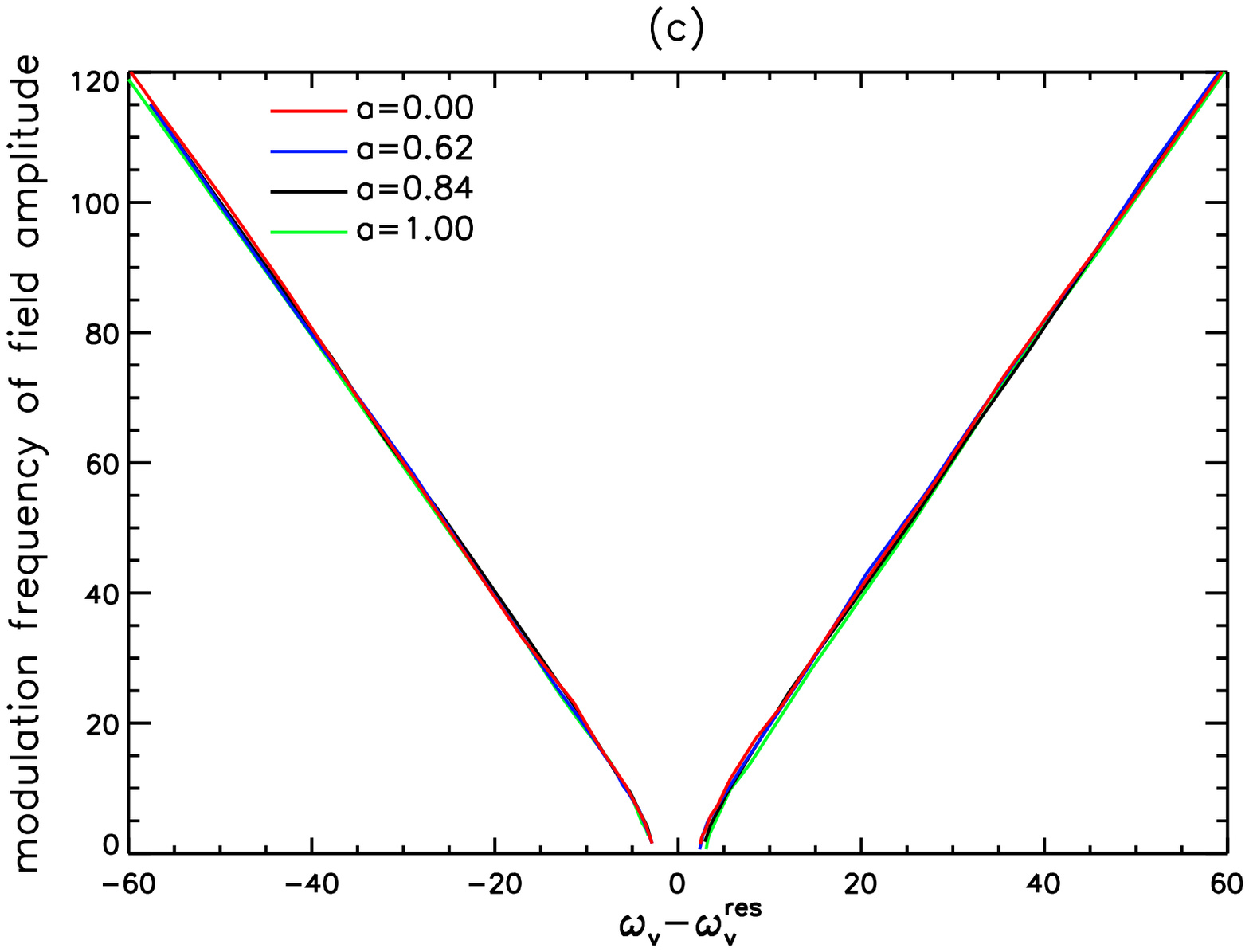}
\caption
{\label{fig::freq_vs_avgdrift} 
(Color online) Panels (a) and (b) show the drift frequency of the
magnetic eigenmode which is computed from the time-derivative of
the azimuthal phase $\partial\varphi/\partial t$.
(a) average field drift frequency
versus the vortex drift frequency for various values of the symmetry
breaking parameter $a$ (value of $a$ increases from top to bottom
curves). The solid curves denote the average field 
drift in the modulated regime. The horizontal dotted lines denote
the field drift in the unperturbed case ($\omega_{\rm{f}}^0$).  (b)
Field drift frequency in the resonant regime.  The field drift is
synchronized with the vortex drift:
$\omega_{\rm{f}}\sim\omega_{\rm{v}}$.  (c) Frequency of
the magnetic field amplitude
modulation for various values of the symmetry breaking parameter
$a$.  The frequencies of the amplitude modulation
scale $\omega_{\rm{a}} \sim  2\omega_{\rm{v}}$.}
\end{figure}

The periodic distortion caused by the drifting vortices also
influences the drift behavior of the magnetic eigenmode.  In the
resonant regime the dynamo eigenmode exhibits an azimuthal field drift
that is immediately locked to the vortex drift frequency, i.e., the
magnetic field pattern follows the vortices [$\omega_{\rm{f}}\sim
\omega_{\rm{v}}$, Fig.~\ref{fig::freq_vs_avgdrift}(b)].
This frequency locking also applies without equatorial symmetry
breaking ($a=0$), which does not show any field drift without vortices
[red (topmost) curve in Fig.~\ref{fig::freq_vs_avgdrift}(b)].  Besides the
temporal decay or growth, the (constant) azimuthal field drift determines
the only time scale in the resonant regime, whereas outside of the
resonant window two different time scales appear: an average azimuthal
field drift that is roughly determined by the equatorial
symmetry-breaking parameter $a$ [solid curves in 
Fig.~\ref{fig::freq_vs_avgdrift}(a)]\footnote{This does not hold in
  the vicinity of the transition between the modulated regime and the
  resonant regime where the field drift exhibits a divergent trend.}
and the aforementioned modulation of the field amplitude which scales
$\sim 2\omega_{\rm{v}}$ [Fig.~\ref{fig::freq_vs_avgdrift}(c)].  In the
modulated regime, the field drift is not constant but varies regularly
with the same timescale as the amplitude modulation.  A movie that
shows the time development of four characteristic runs (no vortex,
resonant run with frequency locking, and modulated case with positive
and/or negative vortex drift frequency) can be found in the
supplemental material for this article \cite{suppmat}.

In order to illustrate the behavior of the eigenmode in the vicinity
of the transition between the resonant and the modulated regime, we
have combined the curves shown in Fig.~\ref{fig::freq_vs_avgdrift}
into one drawing.
\begin{figure}[t!]
\includegraphics[width=8cm]{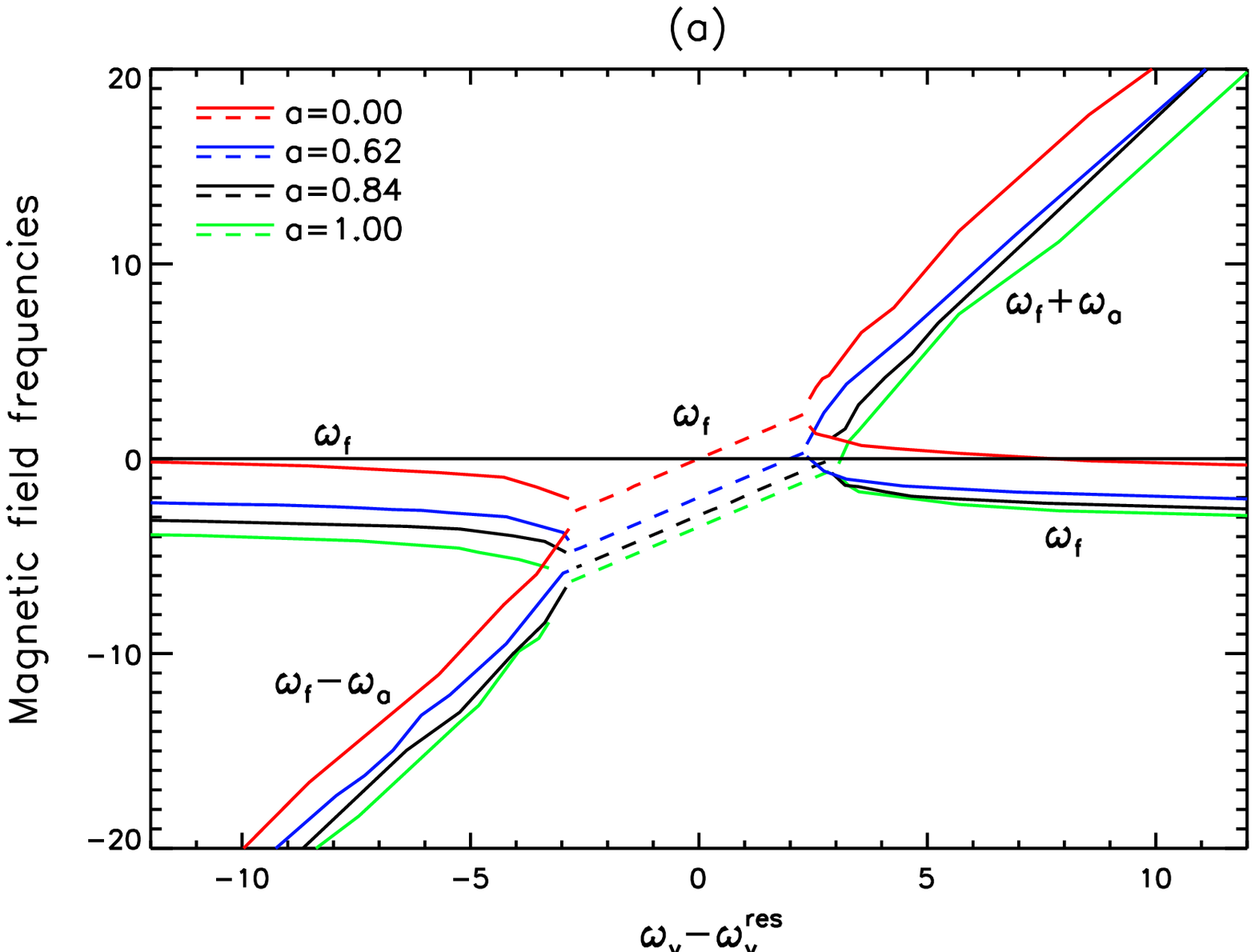}
\includegraphics[width=8cm]{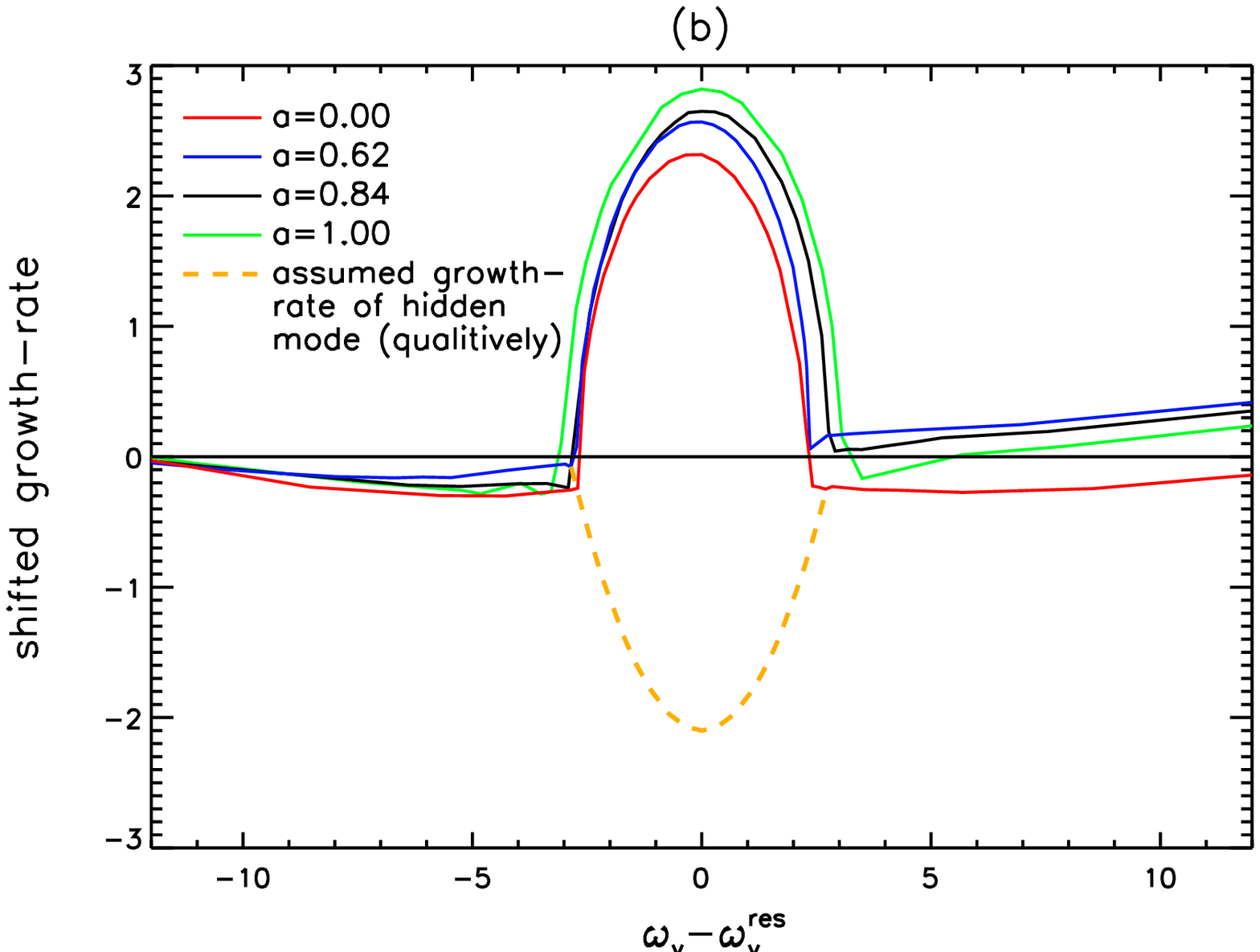}
\caption{(Color online) (a) Compilation of frequencies of the magnetic eigenmode
  versus the vortex drift frequency minus the resonance frequency. The
  solid curves show a combination of field drift frequency and
  amplitude modulation frequency and the dashed central part
  denotes the field drift frequency
  locked to the vortices in the resonant regime.  (b) corresponding
  growth-rates, Note the dashed curve in the bottom part which
  represents the assumed growth-rates of a {\it{hidden}} mode that
  cannot be identified with the applied time-stepping approach.  %
}\label{fig::combined_timescales.ps}
\end{figure}
Figure~\ref{fig::combined_timescales.ps} shows a combination of
frequencies of the magnetic eigenmode
($\omega_{\rm{f}}\pm\omega_{\rm{a}}$ and
$\omega_{\rm{f}}$, left panel)
and the related
growth rates (right panel) against
$\omega_{\rm{v}}-\omega^{\rm{res}}_{\rm{v}}$.  By using the
sum and difference of the observed frequencies (drift and amplitude
modulation), a continuous transition but with a jump in the derivative
is achieved between modulated and resonant regime.  
The abrupt transition from the modulated regime with $\omega_{\rm{a}}
\sim 2\omega_{\rm{v}}$ to an unmodulated regime with $\omega_a=0$ and
$\omega_{\rm{f}}\sim\omega_{\rm{v}}$ occurs when two different
eigenmodes merge.  At these points, known as {\it{exceptional points}}
\cite{katobook}, the eigenvalues of two eigenmodes coincide.  In our
case the change of the temporal behavior of the eigenmode results from
the merging of the imaginary parts of the two eigenvalues, whereas the
real parts (growth rates) are identical in the modulated regime, giving
rise to the field amplitude modulation.  In the resonant regime (where
the field amplitude modulation vanishes) the real parts of the
eigenvalues of both interacting modes split and presumably form a
"bubble" similar to the behavior of periodical perturbed resonant
mechanical systems (see discussion below).  In addition, at the
exceptional point the two previously linearly independent
eigenfunctions collapse and become indistinguishable. Mathematically,
this is described by the formation of a nondiagonal Jordan block
structure in the matrix representation of the (non-self-adjoint) dynamo
operator associated with algebraic eigenvectors
\cite{2004CzJPh..54.1075G}.  A characteristic property of exceptional
points with two coinciding eigenvalues is reflected in the time
dependence of the field amplitude, which, exactly at the exceptional
point, exhibits an additional secular term linear in $t$
\cite{PROP:PROP201200068} so 
\begin{equation}
\vec{B}(t)\sim (\vec{b}_1+\vec{b}_2t)e^{\gamma t}.\label{eq::timebehave_exppoint}
\end{equation}
Such a time dependence has been found experimentally in an examination
of a two-level system in microwave billiards
\cite{2007PhRvE..75b7201D} and it would be interesting to identify
this term in our simulations.  
However, relation~(\ref{eq::timebehave_exppoint}) is 
exactly valid only very close to the transition point between
the modulated and resonant regimes and it seems that, in practice, a
unique decomposition of the "measured" growth rates (obtained from
our time-stepping scheme) according to~(\ref{eq::timebehave_exppoint})
remains impossible. Hence, we can only speculate on the nature of the
interacting eigenmodes because, unfortunately, with our present
time-stepping method, we can only identify the leading eigenmode,
whereas the identification of the second mode would require the use of
an appropriate eigenvalue solver. 
However, it is rather suggestive to assume that the leading mode
essentially consists of a dominant $(m=1)$ component with some slight
addition of a ($m=2$)-vortex-induced $(m=3)$ component and that the
second eigenmode is dominated by a $(m=3)$ component, to which a
slight $(m=2)$-vortex induced $(m=1)$ component is added.
Independently, we expect that the growth rates of this "hidden" mode
qualitatively behave in the same way as in comparable mechanical
systems, forming a bubble in the resonant regime. The assumed
development has been added in
Fig.~\ref{fig::combined_timescales.ps} in terms of the dashed
orange curve.

\begin{figure}[h!]
\includegraphics[width=8cm]{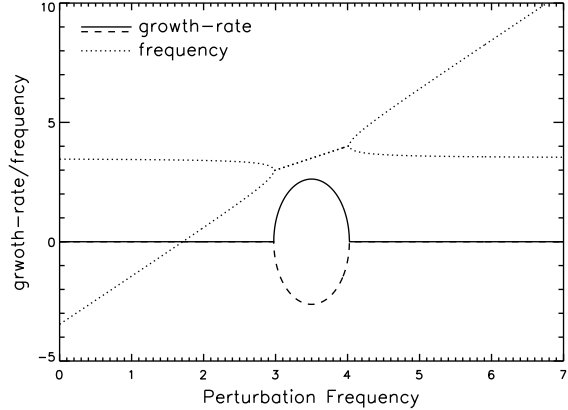}  
\caption{Growth rates and frequencies of an (approximate) solution to
  the Mathieu equation~(\ref{eq::mathieu}) versus the perturbation
  frequency $\widetilde{\omega}$.\label{fig::new}}
\end{figure}

A very similar spectral pattern is well known from mechanical systems subject to
periodic perturbations, which, in the undamped case, can often be
described by a Mathieu-like equation,
\begin{equation}
\ddot{x}+\omega_0^2(1+2\epsilon\cos(\widetilde{\omega}t))x=0,\label{eq::mathieu}
\end{equation}
where $x$ denotes an oscillating state, $\omega_0$ is the
fundamental frequency of the system, and $\epsilon$ is the amplitude of
the periodic perturbation with frequency $\widetilde{\omega}$. 
An approximate solution of Eq.~(\ref{eq::mathieu}) can be given in the form of
\begin{equation}
x(t)=\left(a_0 \cos\left(\frac{\widetilde{\omega}}{2} t\right)+b_0 \sin\left(\frac{\widetilde{\omega}}{2} t\right)\right)e^{\mu t}.
\label{eq::mathieu_solution}\end{equation}
Applying some further simplifications the exponent $\mu$ can be
calculated as
\begin{equation}
\mu\approx\pm\frac{1}{2}\sqrt{(\epsilon\omega_0)^2-(\widetilde{\omega}-2\omega_0)^2}.\label{eq::gr}
\end{equation}
Exponential growth is obtained when the real part of $\mu$ does not
vanish so the unstable regime is restricted to the
interval  $(2-\epsilon)\omega_0 \leq \widetilde{\omega}
\leq (2+\epsilon)\omega_0$. In this regime exponentially growing
solutions occur with a frequency determined by $\widetilde{\omega}/2$.
Outside this resonant regime $\mu$ is
purely imaginary and the frequency of the solution is approximately given by 
\begin{equation}
\omega\approx \frac{1}{2}\widetilde{\omega}
\pm\sqrt{(\widetilde{\omega}-2\omega_0)^2-(\epsilon\omega_0)^2}.\label{eq::freq} 
\end{equation}
Growth rates and frequencies as obtained from~(\ref{eq::gr})
and~(\ref{eq::freq}) are presented in Fig.~\ref{fig::new} and
qualitatively show a good agreement with the pattern obtained in our simulations.  
A comparable pattern is also observed in the eigenvalue spectrum of
oscillating $\alpha^2$ dynamos with nontrivial radial distribution of
$\alpha$ \cite{2003PhRvE..67b7302S,2012_giesecke}.  
However, in those models,
the behavior of the growth rates and the frequencies is exchanged: The
real part of the eigenvalues merge or split (in the same manner as the
frequencies in the left panel of
Fig.~\ref{fig::combined_timescales.ps}) and the imaginary parts
form a bubble corresponding to a restricted regime with oscillatory
(axisymmetric) solutions with two conjugate complex eigenvalues.  

It is further important to state that the simple model given
by~(\ref{eq::mathieu}) and the corresponding
solution~(\ref{eq::mathieu_solution}) can only serve as an analogy for
our three-dimensional problem.
In particular, our model exhibits dynamo action without perturbation
(when ${\rm{Rm}}$ is large enough), whereas the solutions of~(\ref{eq::mathieu}) only show an
{\it{instability}} within a restricted regime. 
Nevertheless, the resulting frequency behavior
denoted by the dotted curves in Fig.~\ref{fig::new} is very
close to the patterns found in our study (see
Fig.~\ref{fig::combined_timescales.ps}).   

Interestingly, in a few simplified cases dynamo problems
were already successfully reduced to a Mathieu-like equation.
For example, the thin disk approximation used in the galactic
$\alpha\omega$-model from Ref. \cite{1990MNRAS.244..714C} allows a
description of the amplitude of the magnetic vector potential by an
equation similar to~(\ref{eq::mathieu}).  A parametric resonance (also
called {\it{swing excitation}}) is observed when the frequency of the
perturbing velocity pattern is twice the natural frequency of the
dynamo.  Likewise, parametric resonances have been found in a
Bullard-type disk dynamo model with a periodic modulation of the disk
rotation. The model is based on a damped variant
of~(\ref{eq::mathieu}) and resonances which allow magnetic field
excitation for rather low velocities of the conducting disk are also
observed at higher frequencies \cite{2010PhLA..374..584P}.  
However, a corresponding reduction of our complicated three-dimensional
dynamo to a simplified model such as in~(\ref{eq::mathieu}) must be left for
future work.

\section{Conclusions}
We have examined kinematic dynamo action of a von K\'arm\'an-like flow
of a conducting fluid in a cylindrical container.  When the flow
breaks the ideal equatorial symmetry of the system, the critical
magnetic Reynolds number for the onset of dynamo action increases with
the amount of symmetry breaking and a time dependence is introduced in
terms of an azimuthal drift motion of the dominant dynamo eigenmode.
The frequency of this drift increases with the amount of symmetry
breaking as well as with the magnetic Reynolds number.

\begin{figure}[t!]
\includegraphics[width=16cm]{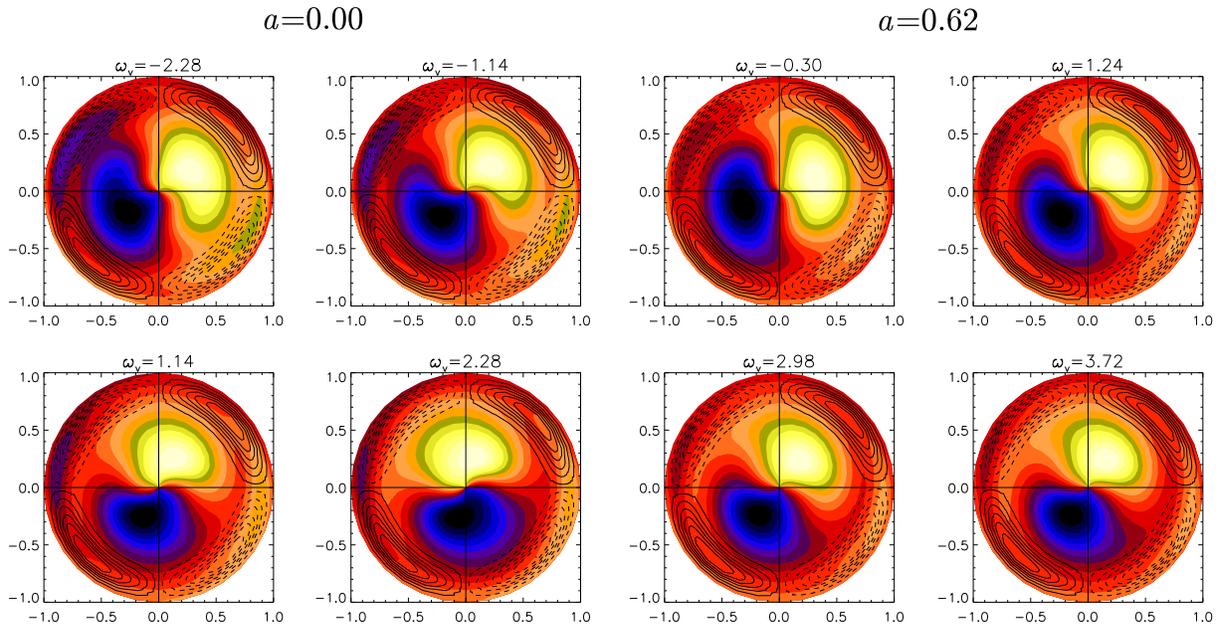}
\caption{(Color online) Alignment of magnetic eigenmode and velocity perturbation for
  different values of the vortex drift frequency
  $\omega_{\rm{v}}$. All runs stem from the resonant regime. The
  color-coded (gray shaded) structure denotes $B_\varphi$ and the contour lines show the
  axial velocity perturbation $u_z^{\rm{v}}$. Left: $a=0$; right:
  $a=0.62$}\label{fig::phase_fieldmode_velmode}
\end{figure}

The main focus of our examinations has been on the interaction of this
field drift with a nonaxisymmetric time-dependent velocity
perturbation and the resulting impact on dynamo action.  In summary,
what we observe is the following: The temporal behavior of the system
is governed by three different time scales: the decay/growth time, the
magnetic field drift and the vortex drift, which, in contrast to the
first two, represents an imposed quantity.  For rather slow vortex
drift frequencies, the first magnetic eigenmode (dominated by an
$(m=1)$ component) is enslaved by the drifting vortex pattern, hence,
we see here $\omega_{\rm{f}}\sim \omega_{\rm{v}}$.  Connected with the
linear frequency relationship in this regime we observe a parabolic
shape of the growth rate, with a maximum close to the resonance point
where the vortex drift frequency would roughly correspond to the
eigenfrequency of the magnetic field for the unperturbed, axisymmetric
flow, and a quadratic reduction of the growth rate nearby this point.
Outside of the resonant regime the field amplitude and the field drift
are modulated with twice the frequency of the vortex drift.
Phenomenologically, the development of growth rates and frequencies can be described by a Mathieu-like
equation. 

The observed behavior can also be explained on the basis of simple
physical principles. For sufficiently slowly drifting vortices the
system adjusts itself to an optimum state and the (azimuthal) phase
between magnetic eigenmode and vortex pattern remains fixed so that
the field growth becomes maximal.  This state is essential
characterized by the alignment between the magnetic eigenmode and the
nonaxisymmetric velocity mode which is roughly the same independently
of symmetry breaking or vortex drift (see
Fig.~\ref{fig::phase_fieldmode_velmode}).  Increasing
$\omega_{\rm{v}}$, the magnetic eigenmode cannot follow the ever
faster $(m=2)$ vortex drift, but "bethinks" of its own eigenvalue (in
the unperturbed state) to which it converges in the limit
$|\omega_{\rm{v}}| \rightarrow \infty$.  In doing so, it will be
"beaten" by the $m=2$ vortex mode with an ever increasing frequency
$2\omega_{\rm{v}}$, which explains the occurrence of the second
frequency involved.

The simulations presented in this study show similarities with the
results from Ref. \cite{2009PhRvE..80e6304R}
where dynamo action was examined using a flow field obtained
from nonlinear simulations of spherical {\it{s2t2}} flow in a sphere.
Kinematic dynamo simulations using the time averaged flow or different
snapshots of the velocity field did not exhibit dynamo action, whereas
this was indeed the case when considering the time-dependent flow.
The exclusive occurrence of dynamo action with a time-dependent flow
was interpreted as dynamo action based on {\it{non-normal growth}}
originally described in Refs. \cite{dormy_gerard-varet_2008,2008PhRvL.100l8501T}. 
Looking at the details, a number of differences become visible between the study of
Ref. \cite{2009PhRvE..80e6304R} and the model presented here. 
The key result of Ref. \cite{2009PhRvE..80e6304R} is a magnetic field
amplification induced by two counterpropagating $m=2$
waves that were anchored at the poles of a sphere. However, in their
study, the authors did not find any clear sharp resonant-like
regime. Furthermore, they only found a more complex temporal
(i.e. cyclic) behavior in one special case with a coupled Navier-Stokes
and induction equation which leads to a phase jump in the alignment between
the nonaxisymmetric velocity mode and the magnetic eigenmode enforced by the
back-reaction of the Lorentz force on the velocity field. 
In all other runs presented in Ref. \cite{2009PhRvE..80e6304R} the
orientation of the magnetic eigenmode remains fixed in space and time
and no azimuthal drift is observed (this behavior is similar to the
results presented in Ref. \cite{2008PhRvL.100l8501T} where only real
eigenvalues of the leading eigenmodes were obtained). 
The most probable explanation for the
distinct behavior results from the differences in the hydrodynamic
base of the model of Ref. \cite{2009PhRvE..80e6304R}, which does not show
any equatorial symmetry breaking (thus, the basic magnetic state is
stationary and non-drifting). Moreover, the wavelike distortions in
Ref. \cite{2009PhRvE..80e6304R} consist
of two counterpropagating $m=2$
patterns that were anchored at the poles of a sphere
which most probably inhibits the phase-locking phenomenon
observed in our study.
As a further difference, we observe a resonant behavior even in runs without
equatorial symmetry breaking and with stationary vortices, i.e., in
systems without any time dependence so no time-dependent
contribution is available that may provide for a mixture of non-normal
modes.  Hence, the role of non-normal growth in our models and the
comparability with the model of Ref. \cite{2009PhRvE..80e6304R}
must for now remain an open question.

It is difficult to conclude if the resonance effect can be realized
in existing dynamo experiments.  Although a coincidence of forcing
frequency (the frequency of the drift motion) and magnetic field
frequency cannot be ruled out, the resonance condition is a quite
particular case and most probably can only be fulfilled by chance.
This is particularly true in the case of a more realistic non-linear
analysis, in which the Navier-Stokes and the induction equation are
coupled so the drift of the vortex pattern might strongly be
influenced by back-reaction of the magnetic field.  Furthermore, in
the examined VKS-like configuration, the vortex drift frequencies
observed in the water experiment are far away from the resonance
condition (at least for reasonable values of the equatorial symmetry
breaking) so it is unlikely that the vortices affect the VKS
dynamo in its actual setup.  Nevertheless, it might be suggestive to
change the large-scale flow geometry in order to adjust the relation
of vortex drift and field drift, e.g., by changing the aspect ratio
\cite{1993Ap&SS.208..245K} or by fixing the vortices using some wall
inhomogeneity.
From Table~\ref{tab::freq} it follows immediately that the most
promising state to realize the resonance 
in the experiment would be in a configuration that suppresses the
equatorial symmetry breaking (e.g., using a ring in the equatorial
plane). In that case, the resonance maximum occurs for nondrifting
vortices ($\omega_{\rm{v}}=0$), providing a reduction of $\sim$15\% (from
${\rm{Rm}}^{\rm{c}}=59.7$ to ${\rm{Rm}}^{\rm{c}}=50.5$). 
This configuration is definitely achievable in the experiment where
the vortices can be anchored by mounting some
inhomogeneity on the outer cylinder wall like, e.g., holes or
fingers. However, even when the resonance condition theoretically is
adjusted, it remains unclear whether a parametric resonance condition
can be fulfilled because in the highly turbulent regime the vortex
position undergoes considerable fluctuations. These
fluctuations (which indeed are observed in the water experiment) could be
modeled by introducing a random phase in the equations for the
nonaxisymmetric flow perturbation but, in contrast to 
amplitude fluctuations, such phase noise is known to prevent the
occurrence of a parametric resonance \cite{petrel_fauve_2006}.

\acknowledgments{A.G. and F.S. acknowledge helpful discussions with Oleg
  Kirillov and financial support from Deutsche Forschungsgemeinschaft
  (DFG) in frame of the Collaborative Research Center (SFB) 609. J.B. is
  grateful for the support of the Spanish government through Contract
  No. FIS2011-24642.  The authors thank B. Knaepen and D. Carati of the
  Universit{\'e} Libre de Bruxelles for inviting them to the Brussels
  Summer program 2009.  The computations were performed using GPUs
  provided by the CUDA Research Center at TU Dresden (now CUDA Center
  of Excellence \cite{ccoe}).  }

\bibliographystyle{apsrev4-1}

\end{document}